# Pic2Spec: Generative Modeling Reconstructs Single Cell Raman Fingerprints from Brightfield Images

Srilakshmi Premachandran[1], Amit Kumar Bhuyan[1], Loza F. Tadesse[1,2,3*]

[1]Department of Mechanical Engineering, MIT, Cambridge, Massachusetts, USA
[2]Ragon Institute of Massachusetts General Hospital, MIT, and Harvard, Cambridge, Massachusetts, USA
[3]Jameel Clinic for AI & Healthcare, MIT, Cambridge, Massachusetts, USA

*Corresponding Authors: L. F. T, lozat@mit.edu, S.P, sri01@mit.edu

**Abstract**

Single-cell molecular characterization remains a critical bottleneck in scalable analysis, constrained by labeling requirements, limited multiplexing capacity, and costly reagents that perturb native cellular physiology. Raman spectroscopy offers a compelling label-free alternative through chemically specific vibrational fingerprints, yet its dependence on long acquisition times and reliance on specialized instrumentation have restricted its deployment in high-throughput settings. Here, we demonstrate that vibrational spectral fingerprints can be computationally reconstructed directly from standard brightfield microscopy images, eliminating the need for spectroscopic hardware. We introduce Pic2Spec, a physics-informed generative deep learning framework that learns a shared latent biochemical representation linking morphological image features to underlying vibrational spectral structure, enabling virtual Raman spectroscopy from routine optical micrographs. Validated across both mammalian and bacterial systems, Pic2Spec generates high-fidelity spectra that faithfully reproduce experimentally measured fingerprints, achieving 98% cosine similarity and Pearson correlations of ~95% across diverse datasets, while quantitatively preserving biochemical peak intensities and population-level distributions. Beyond spectral reconstruction, generated spectra retained functional molecular discrimination in bacterial systems, accurately resolving mutation-driven transgenic states and predicting Green Fluorescent Protein expression with performance approaching true Raman measurements and outperforming conventional image-based analysis by 20%. To our knowledge, this represents the first demonstration of chemically informative virtual molecular fingerprints inferred purely from brightfield contrast, replacing time-intensive measurements on expensive Raman instrumentation with computational inference. By reimagining brightfield microscopy as an inference-enabled molecular profiling platform, Pic2Spec democratizes label-free biochemical phenotyping and removes the hardware and acquisition-time barriers that have long confined vibrational spectroscopy to specialized laboratories. This advance opens a scalable path toward high-throughput molecular analysis for clinical diagnostics, drug screening, and longitudinal cellular monitoring at the accessibility of routine optical microscopy.

## Significance Statement

Molecular spectroscopy has long served as a cornerstone of chemical and biological analysis, but its practical deployment is often constrained by instrument costs exceeding $100K, need for specialized expertise, and inherent throughput limitations. Physics-informed generative AI can fundamentally rewrite this paradigm, reimagining spectroscopy from a hardware-bound sphere into a virtual, scalable computational pipeline. Pic2Spec introduces this new framework to predict Raman spectral fingerprints directly from standard brightfield images. In doing so, it transforms routine microscopes into computational "virtual spectrometers," eliminating the instrumentation barrier entirely. This computation-aided translation enables laboratories worldwide to extract chemically specific molecular profiles from existing microscopy platforms without hardware investment, expertise needs, or spectral acquisition delays. Although Raman spectroscopy has shown broad potential for characterizing cell phenotypes, mapping functional states, detecting pathogens, and informing immunotherapy development, its promise for routine biological, clinical, and resource-limited use has remained unrealized. In settings where conventional Raman remains impractical, access to chemically informative spectroscopic analysis has been limited. By linking ubiquitous imaging platforms with chemically informative readouts, Pic2Spec overcomes these barriers, opening the path to broader deployment and discovery across single-cell biology, cell therapy manufacturing, antimicrobial resistance profiling, clinical diagnostics, and beyond.

## Introduction

Quantitative readout of cellular biomarkers underpins modern diagnostics, drug discovery, and functional cell analysis. Yet current techniques capture only a subset of this molecular complexity and require multimodal measurements to yield holistic, actionable insights into cell state. While fluorescence-based approaches offer molecular specificity, they require exogenous labels, staining reagents, or genetic modifications that perturb cell state, severely limit multiplexing capacity, and add substantial cost and preparation burden [1]. Additionally, flow cytometry enables high-throughput analysis but remains dependent on labeling and predefined markers rather than unbiased molecular discovery [2,3]. Furthermore, transcriptomic and omics methods provide deep molecular resolution but are destructive, time-intensive, costly, and fundamentally incompatible with rapid, live-cell screening in clinical or routine settings [4,5]. Among these, Raman spectroscopy stands out as a powerful solution because it directly probes the inelastic scattering of light to generate chemically-specific vibrational fingerprints of biomolecules. These spectra encode highly multiplexed information on lipids, proteins, nucleic acids, and metabolites within individual cells without exogenous labels, operate in aqueous environments, and preserve cell viability for live-cell analysis [6–8]. Its applications span distinguishing cell phenotypes, resolving functional states, identifying pathogens, and directing immunotherapy development [9–11]. Despite these advancements, translation of Raman spectroscopy into routine clinical or resource-limited settings has not yet been realized.

High-fidelity Raman measurements demand high-intensity laser excitation, long integration times, point-by-point acquisition, and high numerical aperture optics, which severely bottleneck throughput and compromise live-cell analysis [12,13]. These constraints become especially prohibitive in studies that resolve heterogeneity across large populations, including microbial phenotyping and immune cell-state analysis. In smaller cells such as bacteria, weak scattering signals exacerbate this challenge, necessitating longer exposures or higher excitation densities. Moreover, Raman measurements demand precise laser alignment, careful focusing and acquisition control, and specialized preprocessing and spectral interpretation, all of which need domain expertise, constraining widespread adoption. Dedicated spectrometers, sensitive detectors, and environmental vibrational isolation demand instrument costs exceeding $100K. Brightfield (BF) imaging lies at the opposite end of this trade-off. It is universal and relatively inexpensive, requires minimal sample preparation, integrates seamlessly with standard cell culture workflows, supports time-lapse imaging, and readily scales to high-throughput acquisition [14,15]. However, brightfield images primarily reflect refractive index and thickness variations and do not directly reveal molecular composition of cells without additional chemical processing steps. Nonetheless, these complementary characteristics highlight a compelling opportunity. If molecular fingerprints could be inferred directly from conventional brightfield images, one could elevate the throughput and quality of information acquisition by combining the biochemical richness of vibrational spectroscopy with the simplicity, speed, and scalability of light microscopy. The central question is therefore to deduce if there exists an inherent relationship between the morphological information in brightfield images and the spectral fingerprint that can enable a

‘pixel-to-spectrum’ translation at the single-cell level. Such a framework that derives the multiplexed biochemical specificity of spectroscopy from brightfield imaging would eliminate the instrumentation barrier entirely and make a significant diagnostic leap without costly hardware, expertise, or time-intensive measurements.

One plausible approach to address the aforementioned goal is an AI-aided cross-modal generative model. Recent advances in deep generative modeling have highlighted the potential of cross-modal translation through shared latent representations[16–19]. However, predicting complete Raman spectra from brightfield images poses a particularly demanding instance because the two modalities encode fundamentally different physical and biochemical information. In biological systems, this gap is further widened by cellular heterogeneity, subtle visual differences, and experimental variability, requiring models capable of learning stable, generalizable mappings between image-derived features and underlying biochemical state. In pursuit of this phenotype-metabolite linkage, we introduce Pic2Spec, a generative deep learning framework that predicts spectral fingerprints directly from single-cell brightfield images acquired on a standard brightfield microscope. We developed generative architectures that combine the properties of inverse problem, representation learning, and cross-modal inference to uncover the mapping between morphological information and biochemical fingerprint. These architectures include latent-variable models inspired by VAEs and convolutional decoders tailored to spectral structure to identify designs that generalize across spectral regions. Using well-characterized bacterial strains and mammalian cells as complementary model systems, we collected paired brightfield images of individual cells with their co-registered Raman spectra to train and evaluate image-to-spectrum models across distinct cell types, morphologies, and biological contexts. The resulting framework delivered a spectroscopic hardware-free virtual Raman fingerprint with greater than 98% cosine similarity and 94-95% Pearson correlation to experimentally measured Raman, capturing key biochemical bands relevant to cell identification. In bacterial systems, the generated spectra was able to discriminate mutation-driven transgenic states within isogenic populations and predict fluorescent protein expression with 88% accuracy, substantially outperforming image-only baselines by 20%, indicating strong preservation of phenotype-relevant spectral information. These results establish that brightfield microscopy, through computational enrichment, can be transformed into a virtual spectrometer, enabling "visual metabolomics"; a direct recovery of molecular fingerprints from morphological images. This work opens a practical route toward high-throughput molecular diagnostics, screening, and real-time cell phenotyping without the hardware, expertise, or time demands that have previously confined spectroscopy to specialized laboratories.

## Results and Discussion

### Pic2Spec generative framework for brightfield-to-Raman translation

Recent advances in deep learning suggest that cross-modal translation between distinct measurement spaces is feasible. Generative models, particularly variational autoencoders (VAEs), have emerged as powerful tools for learning shared latent representations across modalities[19]. In materials science, multimodal VAEs have been used to predict optical absorption spectra from microstructural images and to generate images conditioned on optical properties [20]. More recently, deep learning architectures have been proposed for end-to-end spectral reconstruction from patterned photonic structures, further illustrating how generative models can learn complex mappings between spatial patterns and spectral responses [21]. However, these studies generally involve physics-coupled domains or specialized input modalities, in which the relationship between structure and spectrum is more directly constrained by the underlying system and often guided by substantial prior knowledge. Existing machine learning approaches in spectroscopy have accelerated tasks such as denoising, unmixing, super-resolution, and phenotypic classification, but they still depend on explicit spectral acquisition and often require large training datasets [22–24]. Directly inferring complete molecular spectra from brightfield images presents a substantially harder problem; visually similar cells may correspond to a distribution of biochemical states rather than a single spectral outcome. Addressing this problem requires models that can capture subtle correlations between image-derived cellular features and biochemical state, represent structured biological variability without collapsing to a deterministic estimate, and remain resilient to experimental variability across shifts in cell type, environment, and microscope settings.

Here, Pic2Spec addresses this with a dual-decoder variational architecture for end-to-end translation from single-cell brightfield images to Raman spectra (Fig. 1). Brightfield images are encoded into the parameters of a Gaussian latent distribution, from which latent samples are drawn using the reparameterization trick. A Kullback-Leibler (KL) divergence penalty constrains the learned posterior towards a standard normal prior, promoting a smooth and generative latent space. From this shared latent representation, two decoders are optimized jointly: one reconstructs the input brightfield image, whereas the other predicts the corresponding Raman spectrum. This architecture is designed to preserve cell-resolved morphological structure while organizing latent features in a form that supports biochemical prediction. The image-reconstruction branch, therefore, acts as a structural constraint, reducing the risk that the encoder learns spectral-predictive features that are weakly grounded in the observed cell image. The total training objective combines image reconstruction loss, spectral reconstruction loss, and KL regularization, which are optimized end-to-end by backpropagation. The spectral supervision is designed to preserve multiple aspects of Raman structure by combining MSE, cosine, derivative, and peak-ratio losses, jointly enforcing intensity accuracy, overall spectral shape, local peak dynamics, and relative peak relationships. At inference, Pic2Spec requires only a standard brightfield image of a single cell to generate its Raman fingerprint, eliminating the need for spectroscopic hardware at the point of use. Although training relies on paired image-spectra measurements, deployment uses brightfield images alone, allowing routine microscopy to serve as a computational proxy for molecular readout in settings represented by the training data.

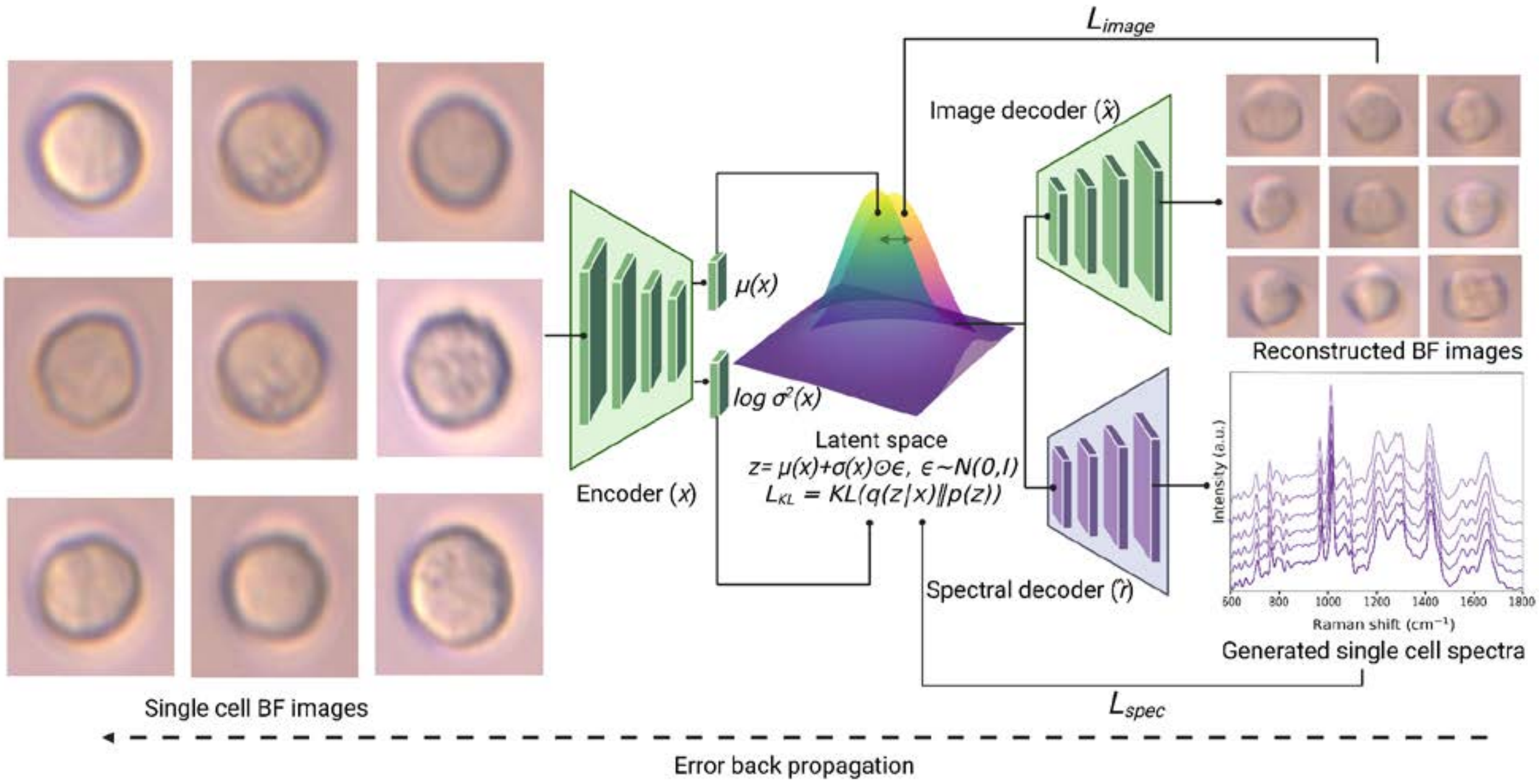

*Fig 1. Schematic representation of Pic2Spec framework: Single-cell BF images are encoded into the parameters of a Gaussian latent distribution, defined by the mean, $\mu(x)$, and log-variance, $\log\sigma^2(x)$. A latent representation, $z$, is sampled using the reparameterization trick with ε ~ N(0,1). A KL divergence term regularizes q(z|x) toward a standard normal prior. The shared latent representation is decoded through two branches: an image decoder producing reconstructed BF images and a spectral decoder producing generated single-cell Raman spectra. Image and spectral reconstruction losses (L_image, L_spec) are jointly optimized with KL regularization via backpropagation.*

**Pic2Spec generates high-fidelity single-cell Raman fingerprints from brightfield images**

To evaluate image-to-spectrum translation in a biologically relevant setting, we established a comprehensive paired dataset consisting of live human Jurkat T cells and primary B cells. They represent biologically distinct immune populations with comparatively subtle morphological differences, making the task non-trivial. The acquired brightfield images were segmented, cropped, and padded at the single-cell level to generate the model inputs. Paired Raman spectra were acquired for the same cells, yielding 206 image-spectrum pairs for T cells and 209 pairs for B cells. Raman spectra were preprocessed using cosmic-ray removal, baseline correction, smoothing, and area-under-the-curve normalization before model training and evaluation. The paired datasets were augmented to capture orientation variability in brightfield images and to better represent spectral broadening and distributional heterogeneity in the Raman data. To ensure rigorous evaluation, data were split into training, validation, and test sets at an 80:10:10 ratio, such that augmented variants of the same cell were confined to a single split. This prevented train-test leakage and ensured evaluation on non-overlapping held-out cells.

Representative brightfield images of T cells are shown, marked with bounding boxes in Fig. 2A. Representative mean generated and measured true spectra are shown in Fig. 2B, with shaded bands indicating ±1 standard deviation across cells. On unseen T cells, Pic2Spec reconstructed Raman spectra with high fidelity across the fingerprint region (600-1800 $cm^{-1}$). Generated spectra reproduce prominent peak structure and maintain qualitative alignment of peak positions with the true spectra, supporting recovery of structured vibrational information. The approach was also further tested with B cells, shown in Fig. 2E. Similar performance was observed for B cells, for which generated spectra closely followed the measured peak shapes and intensity distributions across wavenumbers (Fig 2F). The residual spectra for the above tasks, computed as the difference between the generated and true spectra are shown in Fig.S1. For both T cells and B cells, residuals remained centered near zero across most of the fingerprint region, indicating minimal systematic bias in reconstruction. We next quantified spectral agreement using root mean squared error (RMSE), cosine similarity, Pearson correlation, and spectral angle mapper (SAM). RMSE measures pointwise differences in intensity across the full spectrum, whereas cosine similarity and Pearson correlation assess global agreement in spectral shape and variation. SAM provides a complementary angle-based measure of spectral mismatch, with lower values indicating closer agreement [25]. Across held-out T cells, generated spectra closely matched true measurements, with a median RMSE of 0.00025 [0.00021-0.00034], Pearson correlation of 0.94 [0.90-0.95], cosine similarity of 0.98 [0.97-0.98], and SAM of 11.73° [10.09-14.97°] (Fig. 2C). B-

cell predictions depicted in Fig. 2G showed similarly strong performance, with a median RMSE of 0.000213 [0.000176-0.000259], Pearson correlation of 0.94 [0.90-0.95], cosine similarity of 0.98 [0.97-0.99], and SAM of 10.64° [9.75-13.98°]. Together, these metrics show that Pic2Spec recovers Raman fingerprints with strong global fidelity across both immune-cell datasets.

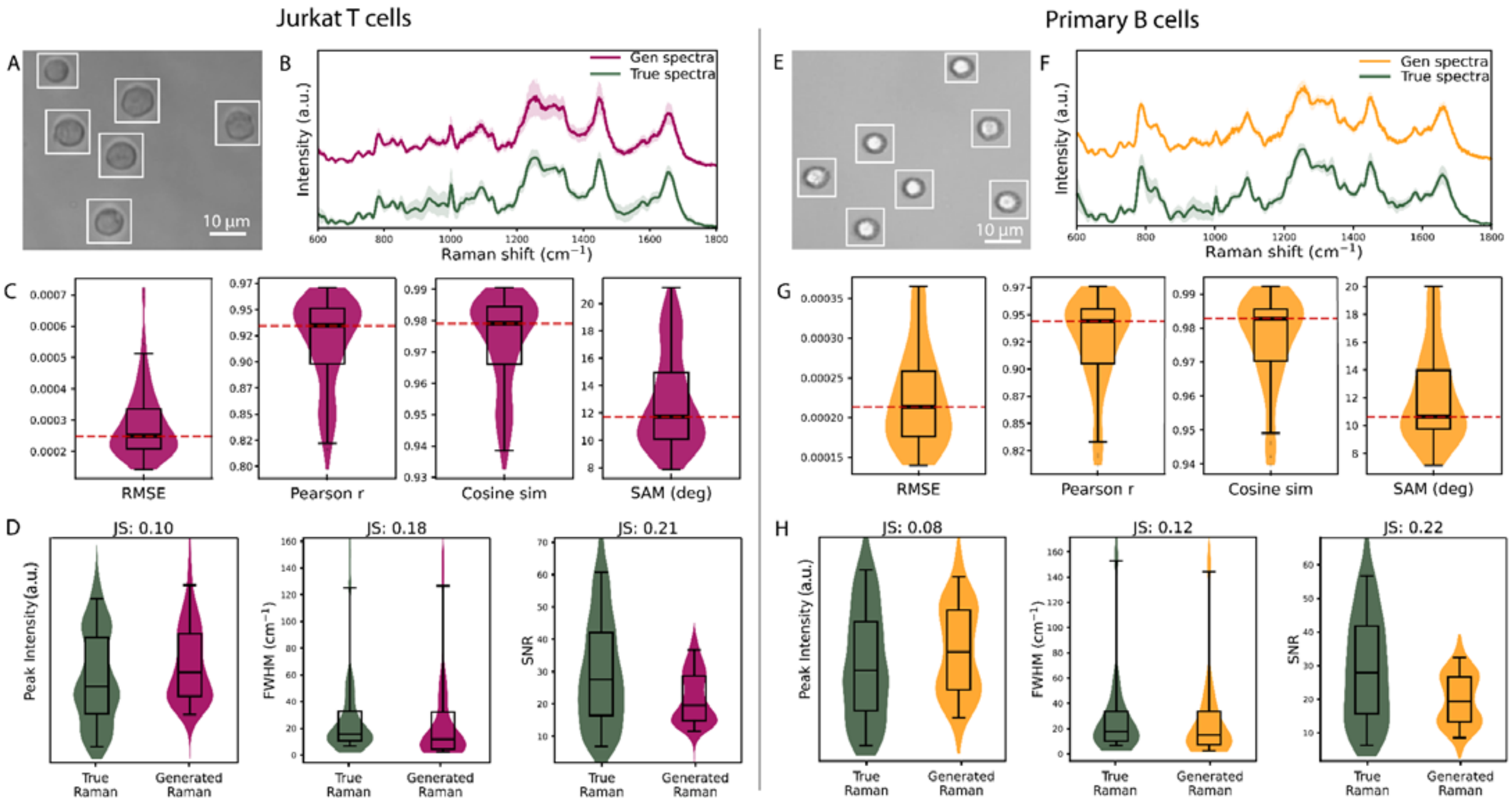

*Fig. 2. Pic2Spec recovers single-cell Raman spectra from brightfield images in immune cells. Left panel: Jurkat T cells. Right panel: Primary B cells. A,E. Representative single-cell BF images used as model input. B,F. Generated Raman spectra with the corresponding measured spectra across the fingerprint region (600-1800 cm$^{-1}$). Shaded regions indicate ± 1 standard deviation (s.d.) C,G. Distributions of spectral similarity metrics on the held-out test set, including RMSE, Pearson correlation coefficient ($r$), cosine similarity, and SAM. Violin plots show the distribution across test cells; boxplots indicate the median and interquartile range. Red dashed lines denote median values (n= 123 T cells, n= 125 B cells) D,H. Comparison of local spectral feature statistics between measured and generated spectra, including peak intensity, FWHM, and SNR for the top 10 prominent Raman peaks. JS divergence values quantify agreement between the distributions of measured and generated spectral features.*

To determine whether this agreement extended beyond summary metrics to individual cells, we examined representative single-cell reconstructions. For T cells, generated spectra showed strong overlap with experimentally measured spectra across major Raman bands (Fig. S4). Generated spectra from individual T cells also exhibited visible variation in peak shape and relative intensity across multiple spectral windows, rather than collapsing to a mean spectrum (Fig. S5). This is evident in the zoomed regions spanning 950-1100 $cm^{-1}$ and 1400-1500 $cm^{-1}$, where cell-to-cell differences were retained in the reconstructions. A similar trend was observed for B cells, where generated spectra closely tracked the true Raman profiles across representative single cells (Fig. S6). The preservation of major peaks and local band structure supports robust single-cell spectral reconstruction in a second immune-cell context. Across B cells, reconstructed spectra shown in Fig. S7 displayed clear inter-cell variability in both the full spectral profiles and zoomed spectral windows between 900-1150 $cm^{-1}$ and 1200-1400 $cm^{-1}$. These findings indicate that Pic2Spec captures cell-resolved spectral structure and relative intensity relationships, rather than merely reproducing population-level averages.

To further assess preservation of local spectral structure, we compared peak intensity, full width at half maximum (FWHM), and signal-to-noise ratio (SNR) for the ten most prominent Raman bands in the generated and measured spectra. For both T cells and B cells, the distributions of peak intensities and linewidths showed strong overlap between predicted and true spectra (Fig. 2D, 2H). This agreement was reflected in consistently low Jensen-Shannon (JS) divergence values: for T cells, JS divergence was 0.10 for peak intensity and 0.18 for linewidth; for B cells, corresponding values were 0.08 and 0.12. By contrast, SNR was modestly downshifted in generated spectra for both cell types, likely because the model learns a denoised, morphology-conditioned estimate of the underlying spectral signal rather than reproducing acquisition-specific measurement noise. As a result, the dominant biochemical features are preserved, whereas stochastic fluctuations and peak-specific intensity extremes may be attenuated. These results demonstrate that Pic2Spec performs accurate cross-domain spectral inference, with the joint latent representation enabling cell-dependent spectra prediction to support computational molecular profiling.

### Image-to-spectrum translation generalizes to bacterial cells and resolves mutation-driven molecular states

Having established that Pic2Spec can recover Raman fingerprints from mammalian-cell brightfield images, we assessed generalizability across different cell types by testing on *Escherichia coli (E. coli)* bacteria, a substantially more demanding regime for single-cell Raman analysis with intrinsically challenging small size and limited scattering volume, yielding weak signals [26]. Across the full bacterial dataset tested, Pic2Spec generated spectra that closely matched the measured Raman spectra over the fingerprint region, with residuals centered near zero and strong agreement in overall spectral shape (Fig. S2(i), (ii)). This agreement was further supported by favorable per-cell similarity metrics and low Jensen-Shannon divergence for peak-level features, indicating preservation of both global spectral fidelity and local peak morphology. Quantitatively, performance on held-out cells remained high, with a median RMSE of 0.000196 [0.000166-0.000245], Pearson correlation of 0.95 [0.92-0.96], cosine similarity of 0.98 [0.97-0.99], and SAM of 10.48° [8.88-13.40°] (Fig. S2(iii)). The distributions of generated and measured spectral intensities also showed close overlap, with low JS divergence for spectral intensity distributions (JS = 0.11) and particularly strong agreement in FWHM and SNR characteristics (JS = 0.04) (Fig. S2(iv)). At the single-cell level, generated spectra reproduced the dominant Raman bands and much of the fine-scale spectral structure observed in experimentally measured spectra (Fig. S8). Agreement across representative cells confirms that the model captures molecularly informative spectral patterns. Generated bacterial spectra varied across individual cells in local band intensity and shape, particularly within diagnostically relevant Raman regions between 750-850 $cm^{-1}$ corresponding to nucleic acids and 1000-1200 $cm^{-1}$ corresponding to proteins, as shown in Fig. S9. These results, along with the mammalian cell results, indicate that Pic2Spec generalizes across distinct cell systems, despite substantial differences in morphology, size, and Raman scattering strength.

Building on this, we further tested whether the observed spectral fidelity holds within phenotypic subgroups beyond the population level. Bacterial heterogeneity often arises within mutation-driven subgroups, and resolving these subpopulations is important for identifying engineered phenotypes, tracking functionally distinct cells within isogenic backgrounds, and detecting early mutation-driven changes[27–29]. We therefore examined whether generative modeling could recover subgroup-specific molecular signals in transgenic bacteria. This provides a stringent test because Raman spectroscopy captures phenotype-relevant molecular differences that are only weakly reflected in brightfield morphology, thereby allowing assessment of whether the generated spectra retain information beyond image-only cues. For this, paired image and spectral datasets were collected across transgenic green fluorescent protein expressing (GFP+) and non-expressing (GFP-) *Escherichia coli* cells. Fluorescence measurements were used solely to assign class labels for downstream classification tasks and were not provided as input to the generative model. Across both phenotypic groups, Raman spectra showed close agreement between generated and true spectra, preserving the overall spectral envelope and major Raman bands, including peaks near 720 and 770 $cm^{-1}$ associated with nucleic acids, 1010 and 1080 $cm^{-1}$ associated with phenylalanine and phosphate vibrations, and the broader 1200-1600 $cm^{-1}$ region corresponding to protein-related modes (Fig. 3A, 3B). For GFP− cells, Pic2Spec achieved a median RMSE of 0.00019 [0.00016-0.00024], Pearson correlation of 0.95 [0.92-0.97], cosine similarity of 0.98 [0.97-0.99], and SAM of 9.99° [8.48-13.28°] (Fig. 3C). GFP+ cells showed similarly strong performance, with a median RMSE of 0.00020 [0.00017-0.00025], Pearson correlation of 0.95 [0.92-0.96], cosine similarity of 0.98 [0.97-0.99], and SAM of 10.89° [9.50-13.45°] (Fig.3D). Additional comparisons of peak intensity, linewidth, and SNR distributions also showed low distributional divergence between generated and measured spectra across both classes (Fig. S3), indicating that spectral fidelity is preserved within individual bacterial phenotypes rather than arising only from pooled population averages.

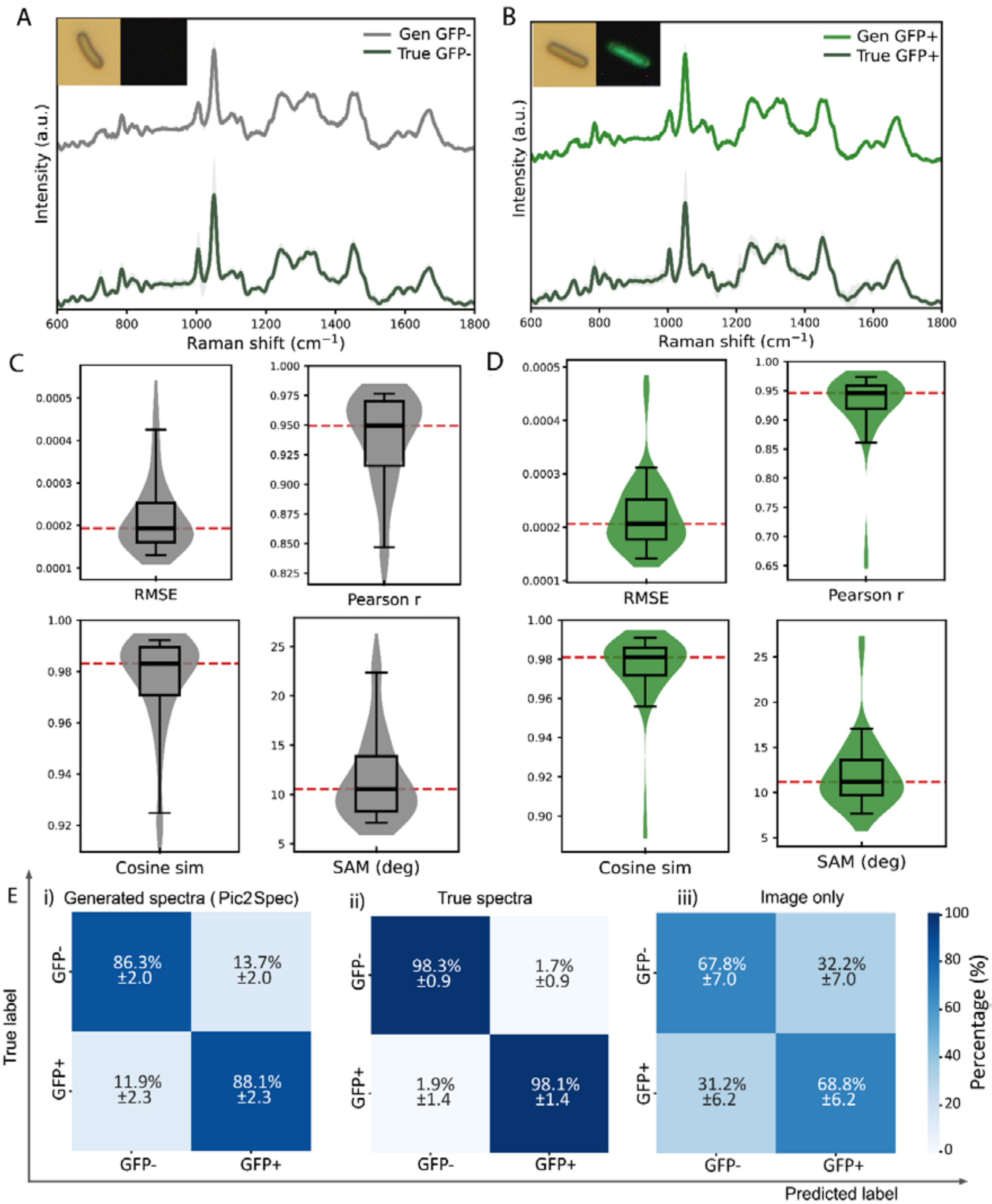

*Fig. 3. Pic2Spec generalizes to bacterial cells and preserves phenotype-discriminative spectral information. Spectral reconstruction performance for GFP− and GFP+ E. coli cells. (A), (B) Generated Raman spectra with the corresponding measured spectra for GFP− and GFP+ cells, respectively. Insets show representative BF and fluorescence images used to assign GFP class labels. Spectra show mean ± 1 s.d. Across n= 132 GFP- cells and n=138 GFP+ cells. C,D, Distributions of spectral similarity metrics on the held-out test set for GFP− and GFP+ cells, respectively, including RMSE, Pearson correlation coefficient ($r$), cosine similarity, and SAM. Violin plots show the distribution across test cells; boxplots indicate the median and interquartile*

*range. For GFP−, n= 132 test cells; for GFP+, n= 138 test cells. E) Downstream phenotype classification using generated spectra, measured spectra, or BF images alone. Confusion matrices are shown for classifiers trained and evaluated using generated spectra from (i) Pic2Spec, (ii) true Raman spectra, and (iii) an image-only baseline. Values in each matrix denote the percentage accuracy ± s.d. of samples assigned to each predicted class.*

We further tested whether the generated spectra retained biologically meaningful discriminatory information by using them for downstream phenotype classification and compared performance against classifiers trained on true spectra and on brightfield images alone. For fair comparison, the same ANN classifier and train-test partitioning were used for true spectra, generated spectra, and image-only baselines. Classification based on generated spectra achieved classification accuracies of 86.3%±2.0% for GFP− and 88.1%±2.3% for GFP+, approaching the performances obtained using true spectra at 98.3%±0.9% and 98.1%±1.4%, respectively (Fig.3E(i), 3E(ii)). By contrast, the image-only baseline achieved an accuracy of 67.8%±7.0% for GFP− and 68.8%±6.2% for GFP+ (Fig. 3E(iii)). These results indicate that Pic2Spec preserves phenotype-discriminative molecular information that is not reliably accessible from brightfield morphology alone, and that the generated spectra retain functional utility for downstream analytical tasks.

**Joint image-spectrum latent learning improves spectral realism**

To determine how architectural design influences image-to-spectrum translation, we compared Pic2Spec with two alternative modeling strategies that impose different assumptions on the mapping from brightfield images to Raman spectra (Scheme S1). The first was a direct cross-modal encoder-decoder (Enc-Dec) that learned a deterministic end-to-end mapping from images to spectra. The second was a latent-translation VAE (Lat-Trans VAE), in which a VAE was first trained to reconstruct brightfield images and the resulting latent representation was subsequently used for spectral prediction[20]. These baselines allowed us to test whether spectral inference is better supported by direct regression, by an image-reconstruction latent space, or by a jointly optimized latent representation explicitly constrained to support both image fidelity and spectral generation.

We evaluated all three architectures, the Dual Dec VAE (Pic2Spec), cross-modal Enc-Dec, and Lat-Trans VAE, across three distinct cell systems spanning *E. coli* bacterial cells, Jurkat T cells, and B cells (Fig. 4A-C). All models were trained and evaluated on the same data partitions and preprocessing pipeline to enable direct architectural comparison. All models were capable of generating spectra that broadly reproduced the measured Raman fingerprint, including major peaks, shoulders, and overall spectral envelope (Fig. 4A(ii), B(ii), and C(ii). Consistent with this, cosine similarity and Pearson correlation remained high across all architectures and cell systems, with the direct encoder-decoder often yielding the highest values on these global similarity metrics (Fig. 4A(iii-iv), B(iii-iv), C(iii-iv)). These results indicate that each model can learn a meaningful mapping from brightfield appearance to spectral output across diverse morphologies and phenotypes.

To further evaluate how these reconstructions behaved at the population level, we examined the distributions and aggregate spectral characteristics of the generated outputs (Fig. 4A-C(ii)). The visual inspection of generated spectra revealed clear differences in how faithfully each architecture preserved population-level spectral structure and variability. In particular, the direct encoder-decoder often generated spectra that appeared over-smoothed and comparatively uniform across cells, consistent with regression-to-the-mean behavior when multiple spectral outcomes are compatible with similar brightfield appearances. The latent-translation VAE generated spectra with more obvious deviations, including attenuated peak amplitudes, broadened features, and departures from the true baseline structure. By contrast, the dual-decoder VAE more faithfully preserved the spectral distribution of the underlying biological data, recovering cell-to-cell variability without collapsing to a single mean-like solution or introducing false fluctuations. This was quantified by analyzing Raman band area and peak height within the biochemically informative 1200-1400 $cm^{-1}$ protein region for T cell spectra, as shown in Fig.4D and 4E, respectively. Although the median values matched with the true spectra, the Enc-Dec architecture markedly compressed the spread of both band area and peak height, consistent with regression toward an average prediction. By contrast, the Dual dec-VAE more closely preserved the variability observed in the true spectra, whereas the Lat-trans-VAE produced broader distributions, indicative of exaggerated spectral noise. Capturing the single-cell variability is especially important for biological applications, where the goal is not merely to reproduce dominant peaks, but to retain the heterogeneity that underlies meaningful cell-state variation across a population.

The advantages of joint latent learning were particularly evident under distributional complexity and more heterogeneous datasets. In bacterial cells, the dual-decoder architecture outperformed the latent-translation VAE by ~3% and maintained stable translation across a mixed dataset containing two phenotypes (Fig.4A(iv)). The inferior performance of the latent-translation VAE suggests that a latent space optimized primarily for image reconstruction is not fully sufficient for downstream molecular prediction, likely because it does not preserve spectral-relevant features as effectively as an architecture trained directly for spectral generation.

The three architectures differ fundamentally in both how the image representation is learned and whether the mapping from brightfield input to Raman output is treated as deterministic or probabilistic. In the direct Enc-Dec, the transformation from image to spectrum is fully deterministic, and the image embedding is optimized solely under spectral supervision. This can support strong task-specific regression and high global similarity, but it also increases the tendency to converge toward smooth average solutions when similar brightfield appearances correspond to multiple plausible spectral outcomes. In the Lat-Trans-VAE, the image representation is learned probabilistically through variational regularization, but this latent space is optimized only for brightfield reconstruction and only subsequently coupled to spectral prediction. As a result, the model captures image variability in a generative manner, yet spectral relevance is not explicitly enforced during latent formation. By contrast, in the Dual dec-VAE(Pic2Spec), the mapping is also probabilistic, but the shared latent representation is shaped simultaneously by brightfield reconstruction and spectral generation. This joint constraint encourages the latent space to encode both structural image content and spectrally informative variation, enabling a more faithful representation of the one-to-many uncertainty inherent in image-to-spectrum translation. The superior spectral realism observed for the dual-decoder architecture is consistent with this balanced, task-aligned probabilistic latent space, which is better suited for cross-domain molecular inference.

We also evaluated a priorless dual-decoder architecture, in which the shared latent representation was learned without variational prior regularization (Fig. S10). In contrast to Pic2Spec with a Gaussian prior, this model failed to generate biologically meaningful Raman spectra, producing outputs with substantial distortion in both overall spectral shape and intensity scaling relative to the measured spectra. The residuals showed large, structured deviations rather than near-zero fluctuations, indicating that the model did not accurately recover the underlying spectral profile. This suggests that without regularization toward a well-structured latent distribution, the shared embedding becomes poorly organized for generative inference, likely limiting its ability to encode spectrally informative features in a smooth and generalizable manner.

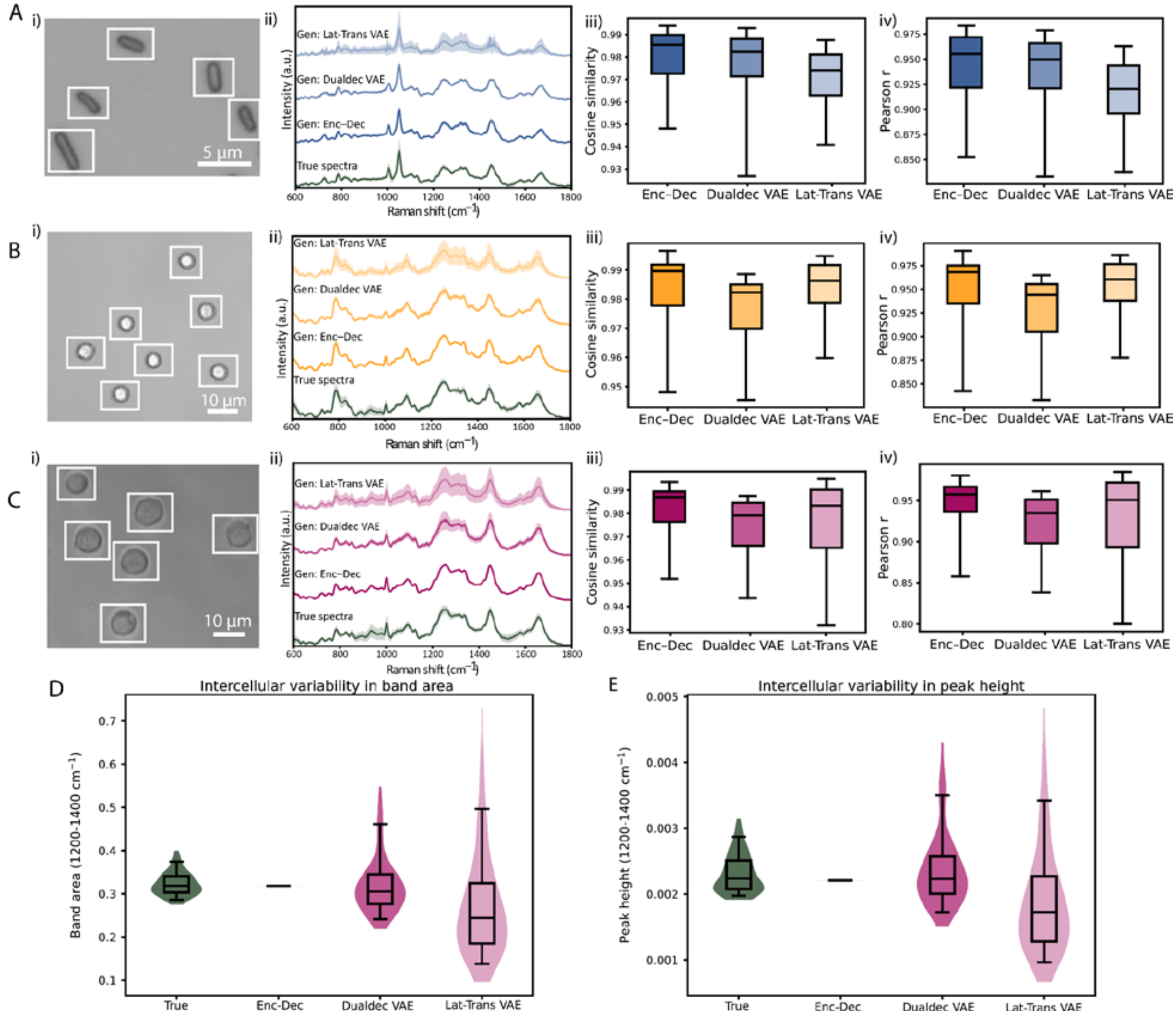

*Fig. 4. Joint image-spectrum latent learning improves spectral realism across distinct cell systems: Comparison of three image-to-spectrum translation architectures: a direct cross-modal encoder-decoder (Enc-Dec), the proposed dual-decoder variational model (Dualdec VAE / Pic2Spec), and a two-step latent-translation variational model (Lat-Trans VAE). A-C, Performance across three cellular systems: A) pooled bacterial cells, B) B cells, C) T cells.(i), Representative BF images from each dataset used as model input. Scale bars: A, 5 µm; B, 10 µm; C, 10 µm. (ii), Raman spectra generated by each architecture stacked with the corresponding true spectrum. (iii) and (iv), Distributions of spectral similarity metrics on the held-out test set, shown as boxplots for cosine similarity and Pearson correlation coefficient, respectively. D and E. Violin plots compare the distributions of band area within the 1200-1400 cm$^{-1}$ region and peak height at 1200-1400 cm$^{-1}$, respectively, for experimentally measured spectra (True) and spectra generated by the Enc-Dec, Dualdec VAE, and Lat-trans VAE models. Box plots indicate the median and interquartile range(n=123).*

**Pic2Spec encodes biochemically informative image features and preserves peak-ratio structure**

To determine whether spectral predictions were grounded in biologically relevant image features, we performed band-targeted saliency analysis by defining scalar targets over three highlighted Raman windows and backpropagating their gradients to the input brightfield image (Fig 5A). We focused on three prominent spectral regions: 700-820 $cm^{-1}$ enriched in nucleic acid associated features, 985-1040 $cm^{-1}$ encompassing phenylalanine and phosphate-related bands, and 1200-1365 $cm^{-1}$ covering amide and CH vibrations linked to protein-rich cell contents. For a representative single bacterial cell, the saliency overlay showed that attribution is concentrated predominantly within the bacterial footprint rather than background regions, indicating that the spectral predictions are driven primarily by cell-associated image features. Notably, this localization is achieved when the model was trained only on paired brightfield and Raman data and did not receive segmentation masks or subcellular annotations. This suggests that Pic2Spec learned to associate morphologically informative cell pixels with downstream spectral output rather than relying on trivial background correlations or acquisition artifacts.

Band-specific saliency maps further revealed that attribution was not spatially uniform across the cell (Fig. 5A). Although all three maps remained largely confined to the bacterial region, their internal hotspot patterns differed across spectral windows, consistent with the idea that distinct parts of the predicted spectrum are supported by distinct image regions. The nucleic-acid-associated and the protein-associated windows each produced localized high-saliency regions within the cell-shaped area, suggesting that the model does not treat the cell as a homogeneous object when forming spectral predictions. Instead, it appears to use localized morphological variations within the brightfield image to support different spectral bands. While saliency maps do not establish direct biochemical correspondence at subcellular resolution, these results suggest that the learned brightfield-to-spectrum mapping is spatially structured and internally selective.

Furthermore, we evaluated whether the generated spectra preserved relative spectral relationships beyond global similarity. To this end, we compared the distributions of selected peak-intensity ratios between true and generated spectra (Fig. 5B). The ratios, $I_{1008}/I_{1120}$, $I_{720}/I_{740}$, and $I_{995}/I_{1010}$, showed close agreement between true and generated distributions, with low JS divergences of 0.17, 0.38, and 0.19, respectively. This indicates that Pic2Spec retained relative intensity structure between functionally related peaks rather than reproducing the average spectral shape. Relatively higher divergence for $I_{720}/I_{740}$ suggests that some local spectral feature relationships are more difficult to recover, potentially reflecting a weaker signal, greater biological variability, or reduced image information for that spectral region. These findings suggest that Pic2Spec does not operate as a purely statistical shape-matching model, but instead learns a structured mapping.

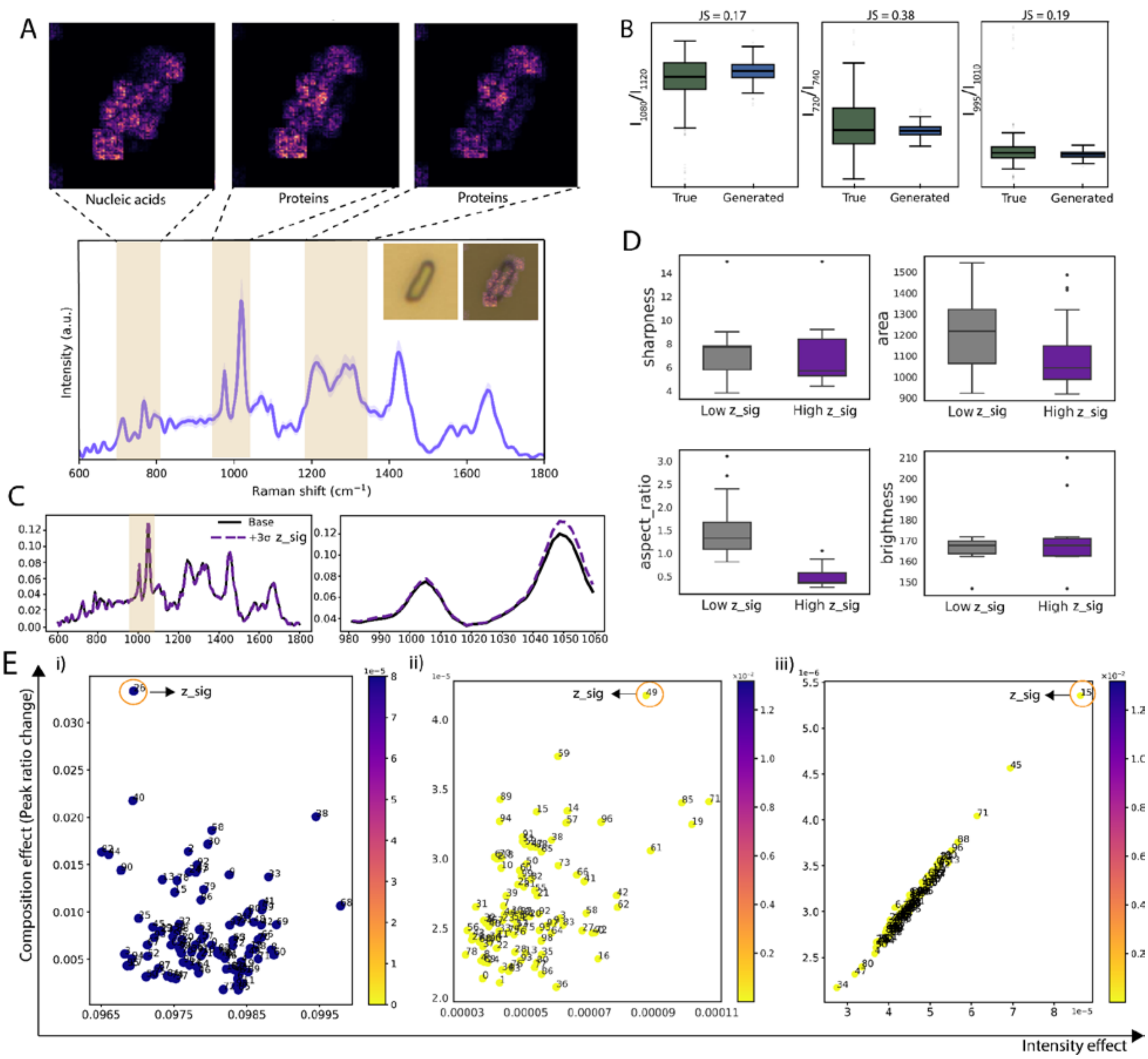

*Fig.5. Band-resolved saliency analysis linking spectral regions to image evidence and ratio-based evaluation of generated spectra. A. Representative spectrum with highlighted Raman windows used for attribution (Inset - BF image of a single bacterial cell and corresponding saliency overlay). Saliency maps for each selected spectral window, annotated by the associated biochemical assignment shown in the figure. B. Distributions of intensity ratios comparing true and generated spectra, with JS divergence reported above each ratio (n=270) . C. Controlled latent dimension perturbations example- the baseline spectrum is compared to the spectrum generated after a +3σ*

*perturbation along a selected composition-sensitive axis. D. Boxplots compare image-derived morphological metrics (projected area and aspect ratio) between low- and high-latent-value groups for a representative composition-sensitive dimension that indicates that latent spectral modulation aligns with measurable morphological variation. E. Two-dimensional summary of all latent axes for (i) Enc-Dec, (ii) Dual dec-VAE, and (iii) Lat-Trans-VAE models, plotted according to their global intensity effect and composition-sensitive (peak-ratio) effect following independent ±σ perturbations.*

**Latent representation disentangles intensity scaling from composition-specific spectral variation**

To examine how Pic2Spec achieves spectral variation, we analyzed the effect of perturbing individual latent dimensions on the generated Raman spectra to study whether the learned representation separates distinct modes of spectral variability. Each latent dimension was perturbed independently while keeping all others fixed, and the resulting spectral response was quantified using two measures: i) global intensity change, defined as the average shift across all wavenumbers, and ii) composition-sensitive change, defined through variations in peak ratios and local spectral structure. When projected into a two-dimensional space (Fig. 5E), latent dimensions exhibit a structured distribution rather than an isotropic spread. The majority of dimensions cluster at low composition effect values (≈0.005-0.015) with relatively narrow intensity variation ($\approx 3\times10^{-5}$ to $5\times10^{-5}$), that indicates dominance of global amplitude scaling. In contrast, a smaller subset of latent dimensions, corresponding to only a few points in Fig. 5E, shows stronger composition effects (≈0.02-0.035), with a broader intensity range (up to $\approx 8\times10^{-5}$), corresponding to dimensions that induce localized spectral modulation. This separation shows that composition-sensitive directions occupy a more confined but distinct region of the latent space. This behavior is also observed in the Dual dec-VAE, where latent dimensions follow a similar distribution, with a clear separation between intensity-dominant and composition-sensitive effects. In contrast, the Lat-Trans-VAE shows a more proportional relationship, where dimensions that induce stronger composition changes also produce correspondingly larger intensity variation, while weaker dimensions remain confined to low values along both axes.

We further evaluated the consistency of these effects using controlled latent dimension perturbations, defined here as systematically increasing or decreasing selected latent coordinates while keeping all remaining latent coordinates fixed. Perturbations along representative composition-sensitive latent dimensions produced reproducible changes in specific Raman bands across multiple test cells (Fig. 5C). These changes are localized and directional, with visible modulation in peak amplitudes while preserving the overall spectral envelope. Despite these local peak-level changes, the perturbed spectra remained highly similar to the baseline generated spectra, with cosine similarity of approximately 0.99. This indicates that latent perturbations selectively alter specific Raman bands without disrupting the overall spectral profile.

To assess whether these composition-sensitive latent variations correspond to measurable morphological features, cells were grouped based on latent dimensions of low and high significance, and distributions of morphological metrics, such as projected cell area, aspect ratio, brightness, and sharpness, were compared (Fig. 5D). Clear shifts are observed in projected cell area and aspect ratio, whereas brightness and sharpness show smaller median changes. Specifically, the image-derived projected cell area, measured from the segmented brightfield cell footprint in pixel-area units, decreases from a median of approximately 1200-1300 in the low-latent group to 1000-1100 in the high-latent group. Furthermore, aspect ratio, defined as a dimensionless shape descriptor computed from the segmented cell footprint, shifts from ≈1.2-1.5 to ≈0.4-0.6, which indicates more elongated versus compact morphologies. This indicates what kind of morphological shapes elevate the values of composition-specific dimensions, which in turn affects the compositional information in the generated spectra. Brightness, measured as the mean grayscale intensity within the cell region, shows a modest increase from ≈165 to ≈170 on the image intensity scale, whereas sharpness, computed as an image-focus metric from local intensity variation, exhibits a slight upward shift from ≈6-7 to ≈7-9. These differences are consistent across samples and indicate that morphological features that are directly related to morphological information, such as aspect ratio and area are more significant in translation whereas features related to image acquisition environment have minimal effect on spectral generation.

Together, these results show that the latent representation learned by Pic2Spec is structured and interpretable. It separates global intensity variation from composition-specific spectral modulation, while maintaining alignment with image-derived morphological features. This organization provides a mechanistic basis for how spectral information is inferred from brightfield images and supports controlled, biologically meaningful spectral generation.

## Conclusions and Outlook

Our results establish a route to computational “virtual Raman” that can turn ubiquitous brightfield microscopes into scalable platforms for molecular inference. By enabling rapid, non-perturbative, single-cell molecular profiling from brightfield images, Pic2Spec offers a practical framework for settings where conventional spectroscopy remains too slow, costly, or operationally restrictive. Because Pic2Spec learns statistical relationships from paired brightfield-Raman training data, its performance depends on the variance, quality, and representativeness of those datasets, and domain shifts arising from changes in microscope optics, sample preparation, or biological state distributions may reduce generalization. Future efforts will therefore prioritize robustness across experimental domains, expanded training diversity, and uncertainty-aware modeling to better define generalization boundaries and accelerate clinical translation. This capability could support high-throughput phenotyping in microbiology, including pathogen identification and antibiotic response screening, non-destructive quality control in cell therapy manufacturing, and longitudinal live-cell studies requiring repeated molecular measurements. The same principle extends to materials characterization, environmental monitoring, and automated discovery workflows, where low-cost imaging could be computationally enriched with spectroscopic information to accelerate screening, characterization, and closed-loop experimentation. Collectively, these advances will consolidate image-to-spectrum translation as a foundational capability in computational molecular profiling, transforming the brightfield microscope from a morphological imaging tool into a scalable, accessible, and chemically informative platform for discovery, diagnostics, and real-time biological analysis across research, clinical, and industrial settings.

## Materials and Methods

Detailed descriptions of cell culture protocols, image and spectral data acquisition, data processing, model architectures, and evaluation metrics are provided in the Supporting Information.

## Data Availability

The source data that support the findings in this work are available from the corresponding author upon reasonable request.

## Code Availability

The code used for spectra generation and classification relies on pre-built Python packages and is available from the corresponding author upon reasonable request.

## Acknowledgements

The authors thank all members of the Tadesse Lab at MIT for providing valuable feedback on the manuscript.

**Author Contributions**

SP: writing (original draft), writing (review and editing), conceptualization, methodology, data curation, investigation, visualization, formal analysis; AKB: writing (original draft), writing (review and editing), formal analysis, validation; LFT: writing (review and editing), conceptualization, validation, resources, supervision, project administration, funding acquisition.

# Supporting Information

## Pic2Spec: Generative Modeling Reconstructs Single Cell Raman Fingerprints from Brightfield Images

Srilakshmi Premachandran[1], Amit Kumar Bhuyan[1], Loza F. Tadesse[1,2,3*]

[1]Department of Mechanical Engineering, MIT, Cambridge, Massachusetts, USA
[2]Ragon Institute of Massachusetts General Hospital, MIT, and Harvard, Cambridge, Massachusetts, USA
[3]Jameel Clinic for AI & Healthcare, MIT, Cambridge, Massachusetts, USA

*Corresponding Authors: Loza F. Tadesse, lozat@mit.edu, S.P, sri01@mit.edu

## Materials and Methods

### 1. Cell culture

E. coli strain Seattle 1946 cells (#25922 and #25922GFP for GFP− and GFP+ cells, respectively) were purchased from American Type Culture Collection (ATCC). Cells stored as frozen glycerol stock were streaked onto polystyrene plates with solidified LB agar media (Teknova, Cat. #L1100). Cells were incubated overnight at 37 ℃. A single bacterial colony was picked up and transferred to 5 mL LB broth (Gibco) using a sterile loop, and then incubated at 37 ℃ with continuous shaking at 250-300 rpm for 6-8 hours. As prepared cells were washed in PBS, resuspended in culture-grade water, and drop-casted on a gold-coated silicon wafer prior to image and spectral data acquisition. B cells used in this study were isolated from a donor leukopak obtained from STEMCELL Technologies. The cells were isolated using the Human B cell isolation kit via magnetic negative selection. The Jurkat T cells were purchased from ATCC (TIB-152). The cells were cultured in complete media consisting of RPMI, 10% Fetal Bovine Serum, and 1% Pen/Strep. Cells were incubated at 37°C with 5% $CO_2$. The cells were centrifuged, resuspended, and imaged on a gold-coated Silicon wafer.

### 2. Acquisition of BF Images and Raman spectra of single bacterial cells

BF images were acquired using a Witec Alpha 300 high-intensity white-light LED lamp for Köhler illumination, providing uniform illumination of the sample. Images of single cells were collected using a 100x objective (Zeiss EC Epiplan-Neofluar, NA = 0.9) at 50% illumination intensity. The Raman spectra were acquired using the same 100x objective lens with a 785 nm laser, at an exposure time of 30 seconds and averaged over two accumulations, with a power of 50 mW for bacterial cells. For immune cells, the spectra were obtained with a 785 nm laser, 50X objective, and exposure time of 15s, averaged over two accumulations. The spectra were collected over the wavenumber range 100-3300 $cm^{-1}$ with a 300/line grating and a center wavelength of 2250 $cm^{-1}$.

### 3. BF image preprocessing

The collected BF images were processed using the Cellpose deep-learning cellular segmentation model to filter single cells morphologically[1]. Each raw image was loaded and smoothed using a Gaussian filter to suppress high-frequency noise while preserving cell boundaries. Segmentation of single cells was performed using the pretrained model type from cellpose (Fig.S11). For each raw image, the model returned a segmentation label mask. To isolate single bacterial cells and exclude clusters, debris, and ambiguous objects, these label masks were post-processed using scikit-image and OpenCV. For each raw image, the corresponding label mask was loaded, and the object-level properties were quantified, including area, solidity, major and minor axis lengths, and perimeter, to set a robust gate size and object-specific thresholds. The minimum and maximum accepted areas were defined as $A_{min}$ = 5th percentile and $A_{max}$=90th percentile, respectively. Objects outside this size range were categorized as clusters or debris and were excluded from single-cell consideration. The raw image, segmentation masks, and bound boxes for cropping single cells are depicted in the Supplementary Figure S11.

For objects within the size window, the aspect ratio was calculated as major axis length/minor axis length, and if AR≥ 1.7, the object was considered rod-like. To ensure that only compact cells were selected, a solidity threshold of ≥ 0.85 was set to exclude irregular or fragmented objects. The resulting single cell labels were combined into a binary mask for cropping cells. Next, single cell images were generated by combining the raw images with the binary masks. For each single cell, the bounding box was marked using the object length and width, and then an additional 12 pixels on all sides. The corresponding region is extracted from the raw image along with the cell index. Border replicate padding was employed to standardize input dimensions to 96x96x3.

## 4. Raman spectral preprocessing

Single-cell Raman spectra were preprocessed using a custom Python pipeline for cosmic ray removal, truncation, baseline correction, smoothing, and normalization. To correct for cosmic ray spikes, each raw spectrum was passed through a 1D median filter, and an intensity vector was computed using a kernel size of 5 points. The difference between the original and median-filtered spectra was calculated at each wavenumber. If the difference exceeded a threshold of 3 intensity units, it was classified as a cosmic ray outlier. For these outlier wavenumbers, the true intensity value was replaced by the median-filtered value at that wavenumber. After cosmic-ray removal, the spectra were truncated to the Raman fingerprint region, 600-1800 $cm^{-1}$, to isolate biologically relevant vibrational modes. Further, baseline correction was performed using the asymmetric least squares algorithm. For each spectrum, the baseline was estimated with the following parameters: smoothness parameter = $1 \times 10^7$, asymmetry parameter = 0.001, and number of iterations = 20. The baseline-corrected spectrum was obtained by subtracting the estimated baseline from true intensities. To reduce noise while preserving peak shape, the baseline-corrected spectra were smoothed using a Savitzky-Golay filter with a second-order polynomial and a window length of seven data points. The spectra were normalized by their total area under the curve across 600-1800 $cm^{-1}$ by numerical integration.

## 5. Data Augmentation

To increase the effective size of training data, the preprocessed single cell images of bacteria were subjected to geometric augmentation using the Pillow imaging library. Each input image was augmented using five geometric transformations, providing invariance to cell orientation. This included horizontal and vertical flips, 90, 180, and 270 degree rotations. Along with the true image, the augmented dataset consists of 6 images per bacterial cell to account for images acquired under varied experimental conditions.

Similarly, corresponding spectral data were augmented to simulate physically motivated experimental variability in calibration, intensity, and noise. For each true spectrum, five augmented spectra were generated by redrawing random parameters for spectral shift (±1.5 $cm^{-1}$), intensity scaling (±10%), additive Gaussian noise (0.5 to 2%), and Gaussian peak broadening (0-3 $cm^{-1}$). The transformations were chosen to mimic realistic variations in wavenumber calibration drift, collection efficiency, noise, and instrument resolution. The original spectra were also retained in the final dataset, resulting in a 6-fold expansion of the dataset.

## 6. Data stratification

The image and spectral dataset consists of data acquired from 2 bacterial classes (E. coli and E. coli GFP+). Data stratification and splitting were performed to ensure class balance and prevent any augmentation leakage between the training, validation, and test data. An 80-10-10 split was performed for training, validation, and test data, respectively. The resulting training set contained 2154 images and corresponding spectra for training, 258 pairs for validation, and 270 pairs for testing. To ensure reproducibility, a random seed of 42 was fixed. The immune cell dataset consisted of 1236 B cell images and spectra pairs and 1254 T cell images and spectra pairs. A similar 80-10-10 split was performed on these datasets for training, validation and testing purposes.

## 7. Model descriptions

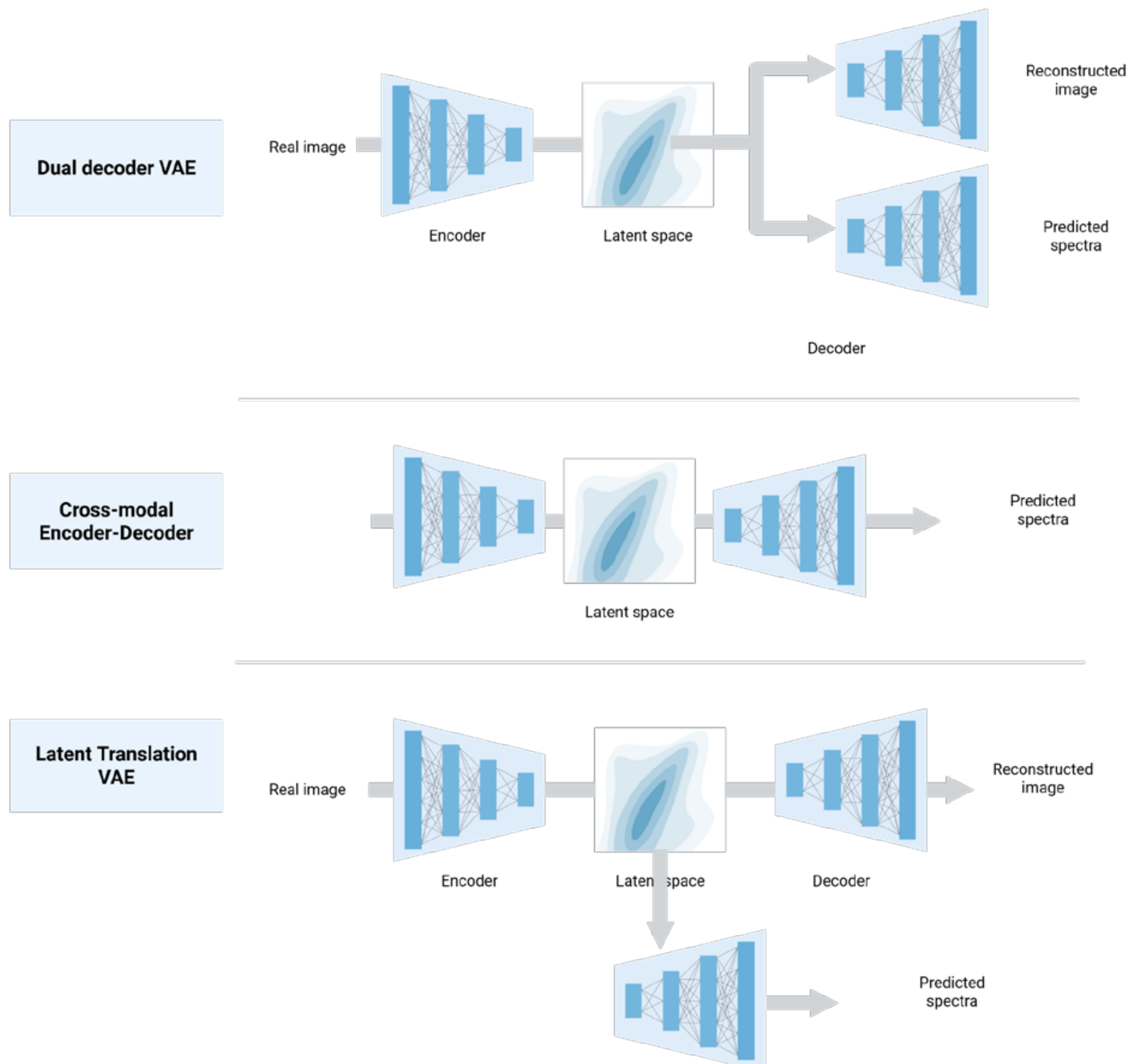

*Scheme S1. Schematic comparison of generative architectures for image-to-Raman translation. The three model designs evaluated in this study are illustrated. Top, Dual-decoder VAE: a shared image encoder maps the input brightfield image into a common latent space that is decoded through two branches to jointly reconstruct the input image and generate the corresponding Raman spectrum. Middle, Cross-modal encoder-decoder: the input image is encoded into a latent representation that is decoded directly into a Raman spectrum without an image-reconstruction branch. Bottom, Latent-translation VAE: an image VAE first learns a latent representation for image reconstruction, after which a separate spectral decoder translates the latent code into a predicted Raman spectrum.*

**Model 1: Dual decoder VAE**

A dual-decoder variational autoencoder (VAE) was developed to learn a shared latent representation from single-cell brightfield (BF) images and to use that latent space for both image reconstruction and Raman spectral generation. VAEs provide a probabilistic latent-variable framework in which an encoder learns an approximate posterior distribution over a low-dimensional latent space and a decoder reconstructs observations from samples drawn from that distribution[2,3]. Here, this framework was extended to a cross-modal setting by coupling a shared image encoder to two decoder branches: an image decoder and a spectral decoder. All models were implemented in TensorFlow 2.19 using the Keras API and trained on an NVIDIA T4 GPU. Input BF images consisted of single-cell images of size 96 × 96 × 3. For an input image $x$, the encoder parameterized a Gaussian approximate posterior over a 100-dimensional latent variable $z$,

$$q_{\phi}(z \mid x) = N(z; \mu(x), log(\sigma^2(x))),$$

where $\mu(x)$ and $\log\sigma^2(x)$ were predicted by two parallel dense layers. Visual features were extracted using four convolutional layers with 8, 16, 4, and 4 filters, respectively, each with 3×3 kernels and ReLU activation. The filter sizes were chosen based on the literature, which showed stable training and validation performance on our dataset[4]. Spatial downsampling was performed using 2×2 max pooling between convolutional blocks, yielding a final encoder feature map of size 6×6×4. This representation was flattened to 144 units and projected to the latent mean and log-variance vectors. Latent sampling was performed using the reparameterization trick,

$$z = \boldsymbol{\mu} + exp(\frac{1}{2}\, log\boldsymbol{\sigma}^2) \odot \boldsymbol{\varepsilon}, \text{ with } \boldsymbol{\varepsilon} \sim N(0,I),$$

which enables gradient-based optimization through the stochastic latent layer.

The image decoder reconstructed cell morphology from $z$. A dense layer first expanded the latent vector to match the final encoder representation (6×6×4), followed by a sequence of upsampling and convolutional blocks with 4, 4, 16 and 8 filters, respectively, each using 3×3 kernels and ReLU activation. A final 3-channel convolutional layer with sigmoid activation produced a reconstructed image $\hat{x}$ of size 96×96×3.

To generate Raman spectra from the same latent representation, a 1D convolutional spectral decoder was designed to progressively upsample along the spectral axis. Each target spectrum was represented as a vector $y \in \mathbb{R}^{573}$, corresponding to the Raman wavenumber grid. The latent vector $z$ was first projected through a dense layer to a sequence of length 72 with 128 channels (72×128 units, ReLU activation). This representation was then passed through three sequential Conv1D-UpSampling1D stages, each consisting of a Conv1D layer with 128 filters, kernel size 5, same padding, and ReLU activation, followed by upsampling by a factor of 2. A refinement block comprising Conv1D layers with 64 filters (kernel size 5) and 32 filters (kernel size 3), both with ReLU activation, was then applied. A final linear Conv1D layer with one filter and kernel size 1 produced the spectral output. Because the upsampling path yielded an output length of 576, the output was truncated to 573 points to match the target Raman grid, yielding the reconstructed spectrum $\hat{y}$.

**Loss formulation**

The model was trained as a multi-output VAE with three objective terms: image reconstruction loss, a custom spectral reconstruction loss, and Kullback-Leibler (KL) divergence between the approximate posterior and a standard normal prior over the latent space. The KL divergence was computed as

$$L_{KL} = D_{KL}(q_{\phi}(z \mid x) \,||\, p(z)) \;=\; -\frac{1}{2}\sum_{j=1}^{d} \quad (1 + log\sigma_j^2 - \mu_j^2 - exp(log\sigma_j^2))\;,$$

where $d$=100 is the latent dimensionality and $p(z)$=$N$(0,$I$).

The image reconstruction term was defined as the mean squared reconstruction error between the input image and reconstructed image,

$$L_{img} = \frac{1}{B}\sum_{i=1}^{B} \quad \frac{1}{CHW} \parallel x_i - \widehat{x_i} \parallel_2^2 \;,$$

where $B$ is the batch size and $C$, $H$, and $W$ denote image channels, height and width. This term was weighted by $\lambda_{img}$=0.5.

The spectral reconstruction loss was designed as a weighted sum of four terms,

$L_{spec} = \lambda_{MSE}L_{MSE} + \lambda_{cos}L_{cos} + \lambda_{deriv}L_{deriv} + \lambda_{ratio}L_{ratio}$.

The first term, $L_{MSE}$, penalized pointwise intensity differences between generated and true spectra,

$L_{MSE} = \frac{1}{B}\sum_{i=1}^{B} \quad \frac{1}{L} \parallel yi - \widehat{y_i} \parallel_2^2$ ,

where $L$=573 is the spectral length.

The second term, $L_{cos}$, preserved overall spectral shape by minimizing cosine distance,

$L_{cos} = \frac{1}{B}\sum_{i=1}^{B} \quad [1 - \frac{y_i \cdot \widehat{y_i}}{\|y_i\|_2 \|\widehat{y_i}\|_2 + \varepsilon}]$,

where $\varepsilon$ is a small constant for numerical stability.

The third term, $L_{deriv}$, encouraged alignment of local peak structure by penalizing differences in the first-order finite differences of the spectra,

$L_{deriv} = \frac{1}{B}\sum_{i=1}^{B} \quad [\frac{1}{L-1} \parallel \Delta y_i - \Delta\widehat{y_i} \parallel_2^2$,

With

$\Delta y_{i,k} = y_{i,k+1} - y_{i,k}$.

This derivative-based term was included to improve preservation of peak positions and local line shapes.

The fourth term, $L_{\text{ratio}}$, constrained selected peak-intensity ratios to preserve relative biochemical relationships across characteristic Raman bands. In general form, for a predefined set of peak pairs $P$,

$$L_{ratio} = \frac{1}{B\mid P\mid}\sum_{i=1}^{B}\sum_{(a,b)\in P}\left(\frac{\widehat{y_{i,a}}}{\widehat{y_{i,b}} + \varepsilon} - \frac{y_{i,a}}{y_{i,b} + \varepsilon}\right)^2.$$

The spectral loss weights were set to $\lambda_{\text{MSE}}$=1.0, $\lambda_{\text{cos}}$=0.2, $\lambda_{\text{deriv}}$=0.05, and $\lambda_{\text{ratio}}$=0.005.

The full training objective was

$L_{\text{total}}=\lambda_{\text{img}}L_{\text{img}}+L_{\text{spec}}+\beta(t)L_{\text{KL}}$,

where $\beta(t)$ is a time-dependent KL weight.

To reduce the risk of early posterior collapse, a $\beta$-annealing strategy analogous to that used in $\beta$-VAEs was adopted[5–7]. Models were trained with a batch size of 64 using the Adam optimizer with an initial learning rate of $3\times10^{-4}$. Training was run for up to 100 epochs with early stopping based on validation loss (patience = 20 epochs) to limit overfitting. The learning rate was adaptively reduced with a factor of 0.5, patience of 8 epochs, and a minimum learning rate of $1\times10^{-6}$. Model parameters corresponding to the lowest validation loss were retained and used for all downstream analyses.

**Model 2: Cross-modal Encoder-Decoder**

The cross-modal encoder-decoder framework predicts Raman spectra directly from BF images using a convolutional image encoder and a 1D spectral decoder implemented in TensorFlow/Keras. Encoder comprises four sequential convolutional layers with 3 × 3 kernels and ReLU activations. The number of filters in the successive convolutional layers was 8, 16, 4, and 4. Each convolution was followed by 2 × 2 max pooling, progressively reducing the spatial resolution from 96 × 96 to 6 × 6. The resulting feature map was flattened and projected to a 100-dimensional latent representation through a fully connected layer. The spectral decoder transformed the latent representation into a Raman spectrum of length 573. The latent vector was first expanded through a dense layer and reshaped into a one-dimensional feature map of length 72 with 128 channels. This intermediate representation was then progressively upsampled through three decoding stages. Each stage consisted of a one-dimensional convolutional layer with 128 filters and kernel size 5, followed by an upsampling operation by a factor of 2. After the three upsampling stages, the decoder output reached a length of 576, which was then refined using two additional one-dimensional convolutional layers with 64 and 32 filters and kernel sizes of 5 and 3, respectively. A final linear 1 × 1 convolution generated the spectral output, which was subsequently trimmed to the target length of 573 channels.The complete model was trained end-to-end by directly coupling the image encoder and spectral decoder, enabling direct translation of BF images into predicted Raman spectra. Model optimization was based on a composite spectral loss. The composite spectral loss used here is the same as the dual decoder model. The model was trained with a batch size of 64 using the Adam optimizer. Training was performed in three sequential phases, during which the derivative-loss weight was adjusted from 0.10 in the initial warm-up phase to 0.05 in the subsequent main and polishing phases to promote early recovery of local spectral structure while maintaining stable convergence. Training used a fixed random seed of 42 for reproducibility.

**Model 3: Latent translation VAE**

To assess whether Raman spectral information could be recovered through an intermediate image-derived latent representation, we implemented a two-stage latent translation framework[4]. In the first stage, a variational autoencoder (VAE) was trained on BF images to learn a compact latent representation of cellular morphology. In the second stage, the learned latent vectors were used as inputs to a multilayer perceptron that predicted the corresponding Raman spectra.

The VAE was built with the encoder taking BF images as input and an image decoder performing reconstruction. The VAE was optimized using a loss comprising an image reconstruction term and a Kullback–Leibler (KL) divergence regularization term. Reconstruction loss was computed as the summed squared difference between input and reconstructed images, and the KL term regularized the approximate posterior toward a standard normal prior. The total VAE loss was defined as the sum of reconstruction and KL losses. Training was performed using the Adam optimizer with a learning rate of $3 \times 10^{-4}$. After VAE training, latent representations were extracted for all images in the training and validation sets using the encoder. For each image, the sampled 100-dimensional latent vector $z$ was used as the representation for downstream spectral prediction. These latent vectors were exported together with image filenames for traceable pairing with their corresponding spectra.

A fully connected regression model was then trained to map the image-derived latent vectors to Raman spectra of length 573. The network accepted a 100-dimensional latent vector as input and comprised three hidden layers with 512, 512, and 256 units, respectively. Each dense layer used swish activation, L2 kernel regularization ($1\text{x}10^{-5}$), and dropout with a rate of 0.10. The output layer generated a 573-dimensional spectrum. A composite spectrum loss was employed, similar to the other models, with the MSE term weighted at 1.0, cosine term at 0.4, derivative term at 0.25, and peak ratio term at 0.005. The network was trained for up to 100 epochs with a batch size of 64 using the Adam optimizer with a learning rate of $3 \times 10^{-4}$.

## 8. Evaluation metrics

Model performance was evaluated by comparing each predicted Raman spectrum variable($\hat{y}$) with its corresponding ground-truth spectrum variable(y) using four complementary metrics: root mean squared error (RMSE), cosine similarity, Pearson correlation coefficient, and spectral angle mapper (SAM). All metrics were computed independently for each spectrum over the full spectral length. The metrics were calculated as follows:
The root mean squared error (RMSE) was defined as

$$RMSE(y,\hat{y}) = \sqrt{\frac{1}{n}\sum_{i=1}^{n} (y_i - \hat{y}_i)}^{\,2}$$

Cosine similarity was computed as:

$$Cosine\ Similarity\ (y,\hat{y}) = \frac{\sum_{i=1}^{n} y_i\hat{y}_i}{\sqrt{\sum_{i=1}^{n} {y_i}^2}\sqrt{\sum_{i=1}^{n} {\hat{y}_i}^2}}$$

Pearson correlation coefficient was calculated as:

$$r(y,\hat{y}) = \frac{\sum_{i=1}^{n} (y_i - \underline{y})(\hat{y}_i - \underline{\hat{y}})}{\sqrt{\sum_{i=1}^{n} (y_i - \underline{y})^2}\sqrt{\sum_{i=1}^{n} (\hat{y}_i - \underline{\hat{y}})^2}}$$

Where $\underline{y}$ and $\underline{\hat{y}}$ are the mean intensities of true and predicted spectra, respectively.

Spectral Angle Mapper was defined as:

$$SAM(y,\hat{y}) = cos^{-1}\left(\frac{\sum_{i=1}^{n} y_i\hat{y}_i}{\sqrt{\sum_{i=1}^{n} {y_i}^2}\sqrt{\sum_{i=1}^{n} {\hat{y}_i}^2}}\right)$$

Where n is the number of spectral points(wavenumbers).

## 9. Peak characteristics analysis

Peak-level spectral characteristics were quantified from each Raman spectrum to assess whether generated spectra preserved key local features of the true measurements. For each spectrum, peak detection was performed within the Raman window using a prominence-based approach implemented in SciPy.

Peaks were required to exceed a minimum prominence threshold defined relative to the dynamic range of the spectrum, $p_{min} = 0.03 \text{ x } (max(y) - min(y))$, where y is intensity. To avoid redundant detection of closely spaced local maxima, a minimum peak separation of 6 cm$^{-1}$ was enforced. The 10 most prominent peaks were retained for all analyses. For each peak, three properties were calculated: peak intensity, full width at half maximum (FWHM), and signal-to-noise ratio (SNR). Peak intensity was defined as the spectral intensity at the detected local maximum. FWHM was estimated at a relative height of 0.5. SNR was calculated by dividing the peak intensity by a robust estimate of the spectral noise. Noise was estimated from the first-difference signal, $\Delta y_i = y_{i+1} - y_i$, which reduces sensitivity to low-frequency baseline structure. The noise standard deviation was approximated from the median absolute deviation (MAD) of the first differences as $\sigma = \frac{MAD(\Delta y)}{0.6745\sqrt{2}}$,

And SNR was computed as $SNR = \frac{h}{\sigma}$, where h is the peak intensity.

Peak properties were extracted independently from each true and generated spectrum, and the resulting values were pooled across all spectra within each group to compare their global distributions. Violin plots were generated using kernel density estimation with Silverman bandwidth selection, whereas the accompanying box plots summarized the median and interquartile structure. To further quantify similarity between true and generated distributions, the Jensen-Shannon (JS) divergence was computed for each feature using histogram-based estimates over a shared value range. Lower JS divergence indicates closer agreement between the peak-property distributions of generated and experimentally measured spectra.

### 10. Saliency Mapping

Gradient-based saliency maps were generated to identify image regions contributing to spectral prediction. For a given input BF image $x$, the trained model produced a predicted spectrum $\hat{y}$, and a scalar target $s(\hat{y})$ was defined as the summed predicted intensity within a selected spectral band. Band limits specified in wavenumber space were mapped to the corresponding spectral channel indices. Saliency was computed as the absolute gradient of the target scalar with respect to the input image, $S = \left|\frac{\partial s(\hat{y})}{\partial x}\right|$. For visualization, gradients were averaged across image channels to obtain a two-dimensional saliency map and min–max normalized to the range [0,1].

### 11. Classification models

Binary classification of GFP+ and GFP− *E. coli* was performed to determine whether generated Raman spectra preserved class-relevant molecular information. A common held-out test set was maintained across all evaluation settings to ensure direct comparability. Three matched testing conditions were used: Test 1, classification from experimentally acquired true Raman spectra; Test 2, classification from generated Raman spectra for the same test samples; and Test 3, classification from brightfield images alone. Evaluation was repeated across 10 independent data splits, and the mean accuracy and standard deviation were reported.

For spectra-based classification, a one-dimensional convolutional neural network (1D-CNN) was used to learn discriminative spectral patterns directly from the Raman profiles. The network consisted of three sequential convolutional blocks with 16, 32, and 64 filters, kernel sizes of 7, 5 and 3, respectively, each followed by batch normalization and max pooling in the first two blocks. The final convolutional representation was aggregated using global average pooling, followed by a fully connected layer with 64 units and ReLU activation, dropout (0.30), and a sigmoid output node for binary classification. The model was trained using the Adam optimizer (LR $1 \times 10^{-3}$) with binary cross-entropy loss, and classification accuracy was used as the performance metric.

For image-based classification, brightfield images were classified using a two-dimensional convolutional neural network (2D-CNN). The architecture comprised three convolutional blocks with 16, 32 and 64 filters, each using 3×3 kernels and ReLU activation, followed by batch normalization and max pooling. Feature maps were globally pooled and passed through a dense layer with 64 units and ReLU activation, followed by dropout (0.30) and a sigmoid output layer for binary prediction. The image classifier was trained using the Adam optimizer (LR $1 \times 10^{-3}$) and binary cross-entropy loss.

### 12. Interpretability Analyses

To interrogate how the learned representation organizes spectral variability and to determine whether individual latent variables encode physically meaningful transformations, systematic perturbation-based analysis of the latent space was performed. This approach examines how controlled perturbations of latent coordinates propagate through the spectral decoder and alter the generated Raman spectra.

Let $z \in \mathbb{R}^d$ denote the latent representation inferred from a BF image through the encoder, and let $\hat{r} = f_\theta(z)$ represent the Raman spectrum generated by the spectral decoder $f_\theta$. For each latent dimension $k$, perturbations were introduced by shifting that coordinate while holding all other coordinates fixed, $z^{(k,+)} = z + \sigma_k e_k$, where $e_k$ is the unit vector along dimension $k$, and $\sigma_k$ is the empirical standard deviation of that dimension computed across the test latent distribution. This perturbation produces a modified spectrum $\hat{r}^{(k,+)} = f_\theta(z^{(k,+)})$ that allows the spectral influence of each latent axis to be quantified. Two complementary metrics were used to characterize the effects of perturbations. The first measures global intensity modulation, defined as the mean change in spectral magnitude using the $L2$ norm, $I_k = \mathbb{E}[|(||\hat{r}^{(k,+)}||_2 - ||\hat{r}||_2)|]$ where the expectation is taken over all test samples. This quantity captures whether a latent dimension primarily acts as an amplitude scaling factor that uniformly increases or decreases spectral intensity. The second metric measures composition-sensitive spectral modulation through peak-ratio changes. For a representative Raman band centered near 1020 $cm^{-1}$, the peak ratio was defined as $\rho = r_{1020}/\Sigma_i r_i$ where $r_{1020}$ denotes the spectral intensity at the selected wavenumber and $\Sigma_i r_i$ represents total spectral intensity. The compositional sensitivity of latent dimension $k$ was quantified as $C_k = \mathbb{E}[|\rho^{(k,+)} - \rho|]$ which reflects perturbation-induced changes in relative peak structure independent of global scaling. Mapping all latent axes into the two-dimensional space $(I_k, C_k)$ reveals a structured organization of spectral variability. A subset of latent dimensions primarily modulated global spectral amplitude while leaving peak ratios largely unchanged which indicates amplitude-dominant directions. In contrast, other axes induced substantial peak-ratio variation with comparatively small changes in total intensity. These dimensions therefore encode compositional spectral modulation rather than simple scaling. The resulting separation indicates that the learned representation partially disentangles global intensity variability from band-specific biochemical structure. To further isolate composition-sensitive behavior, spectral responses were examined after removing amplitude scaling effects. Generated spectra were first normalized by their $L2$ norm, $\check{r} = r/||r||_2$ so that perturbation-induced changes reflected spectral shape rather than magnitude. For selected latent dimensions with large composition sensitivity, the differential spectral response $\Delta r_k = \mathbb{E}[\check{r}^{(k,+)} - \check{r}]$ was computed across the test dataset. These shape-only responses revealed localized modulation of narrow spectral bands rather than uniform vertical shifts. Perturbations consistently altered structured features within the Raman fingerprint region which demonstrates that specific latent coordinates encode selective biochemical contrasts. To evaluate whether these latent directions induce stable transformations across samples, controlled traversal experiments were conducted. For representative cells, spectra generated from the baseline latent code were compared with spectra generated after a $+3\sigma_k$ perturbation along selected composition-sensitive dimensions. Across multiple independent cells, traversal along the same latent axis produced reproducible modulation of the same spectral bands while preserving the overall spectral envelope. The consistency of these responses indicates that the axes correspond to systematic transformations rather than sample-specific noise. Finally, to determine whether composition-sensitive latent variation corresponds to measurable morphological differences in the input images, we examined associations between latent values and image-derived features. For representative dimensions, samples were stratified into low and high latent-value groups and compared across several morphological descriptors computed from the segmented cell images. These included projected cell area, aspect ratio, brightness statistics, and image sharpness. Systematic differences in these metrics between latent groups suggest that spectral modulation encoded in the latent space correlates with measurable morphological characteristics of the cells.

**Supporting Results**

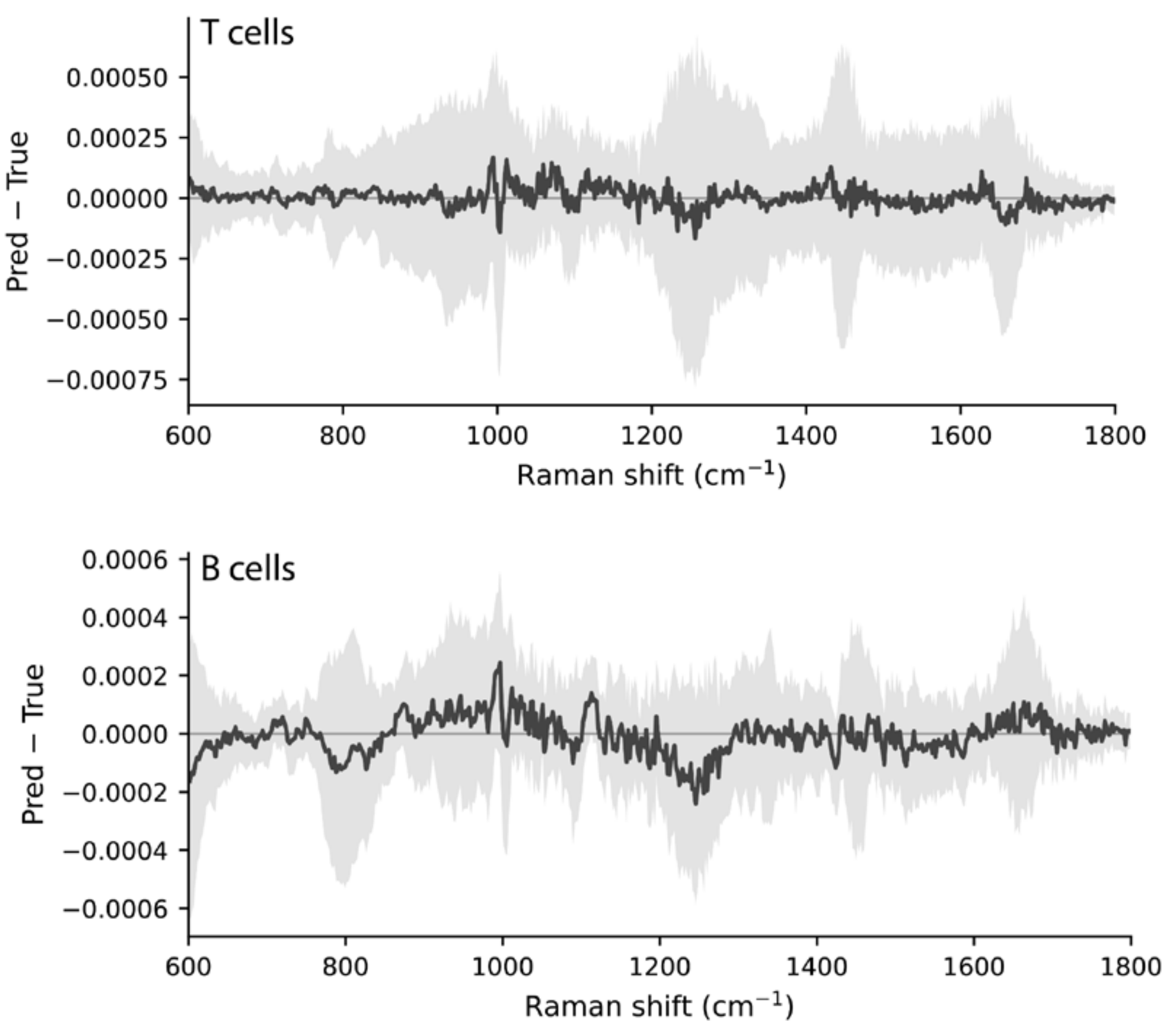

*Fig. S1. Residual spectral error between predicted and true Raman spectra in T cells and B cells. Mean residual spectra, calculated as predicted minus true Raman intensity for T cells (top) and B cells (bottom). Plots indicate the mean residual across (n = 123 T cells, n=125 B cells) single-cell spectra, and the grey shaded region denotes ±1 s.d. Residuals remain centered near zero across most of the spectral window, indicating limited systematic bias in spectral prediction.*

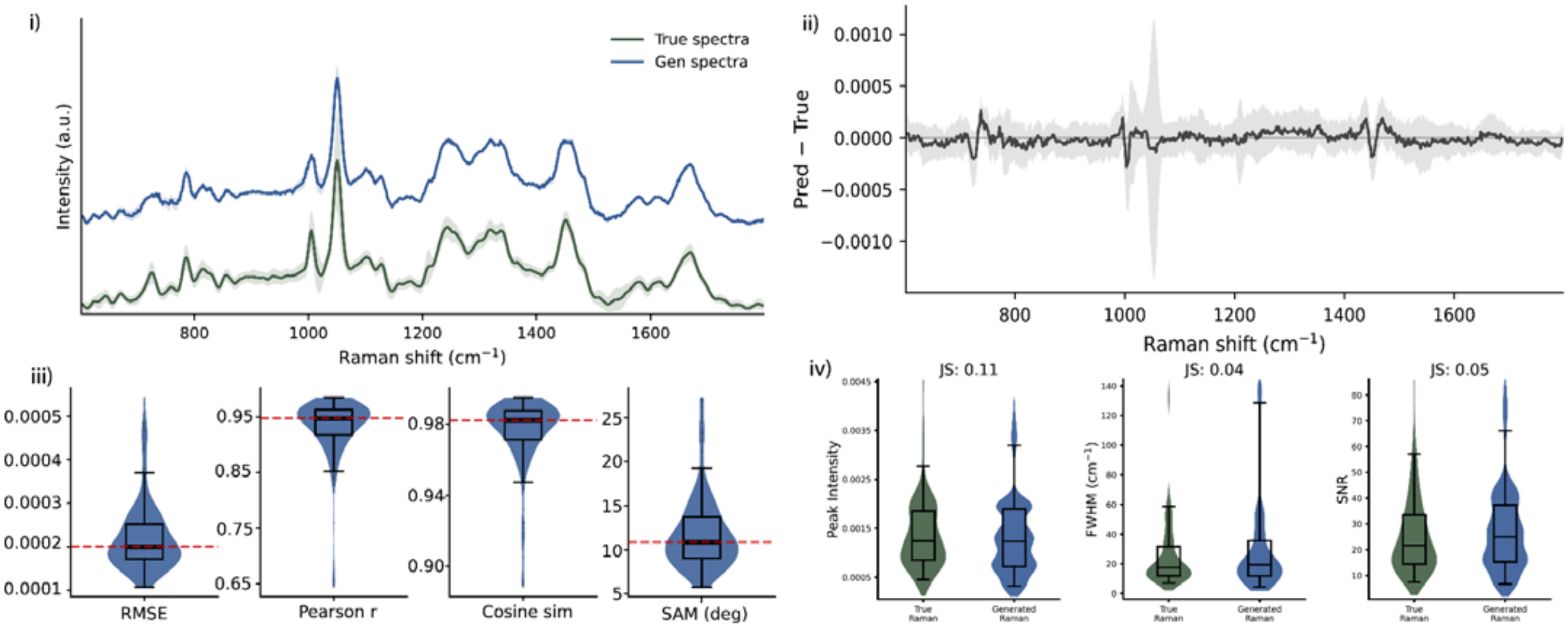

*Fig. S2. Reconstruction fidelity of generated bacterial Raman spectra. i) Representative mean Raman spectra for the bacterial dataset, comparing experimentally acquired true spectra (green) and model-generated spectra (blue). Spectra are shown as mean ± s.d. over n = 270 cells. ii) Residual spectrum calculated as Pred - True across the same wavenumber range. The solid line indicates the mean residual, and the shaded band denotes ±1 s.d. across the test set. iii) Distribution of per-spectra similarity metrics comparing generated and true spectra, including RMSE, Pearson correlation coefficient, cosine similarity, and SAM. Violin plots show the full distribution across test cells, with embedded box plots indicating median and interquartile range. Red dashed lines indicate the median. iv) Distributional comparison of extracted spectral peak characteristics between true and generated spectra, including peak intensity, FWHM, and SNR. Violin plots show pooled peak-level measurements across n = 10 peaks, and JS divergence values quantify agreement between the corresponding true and generated distributions. Lower JS values indicate closer correspondence in local peak morphology and spectral quality.*

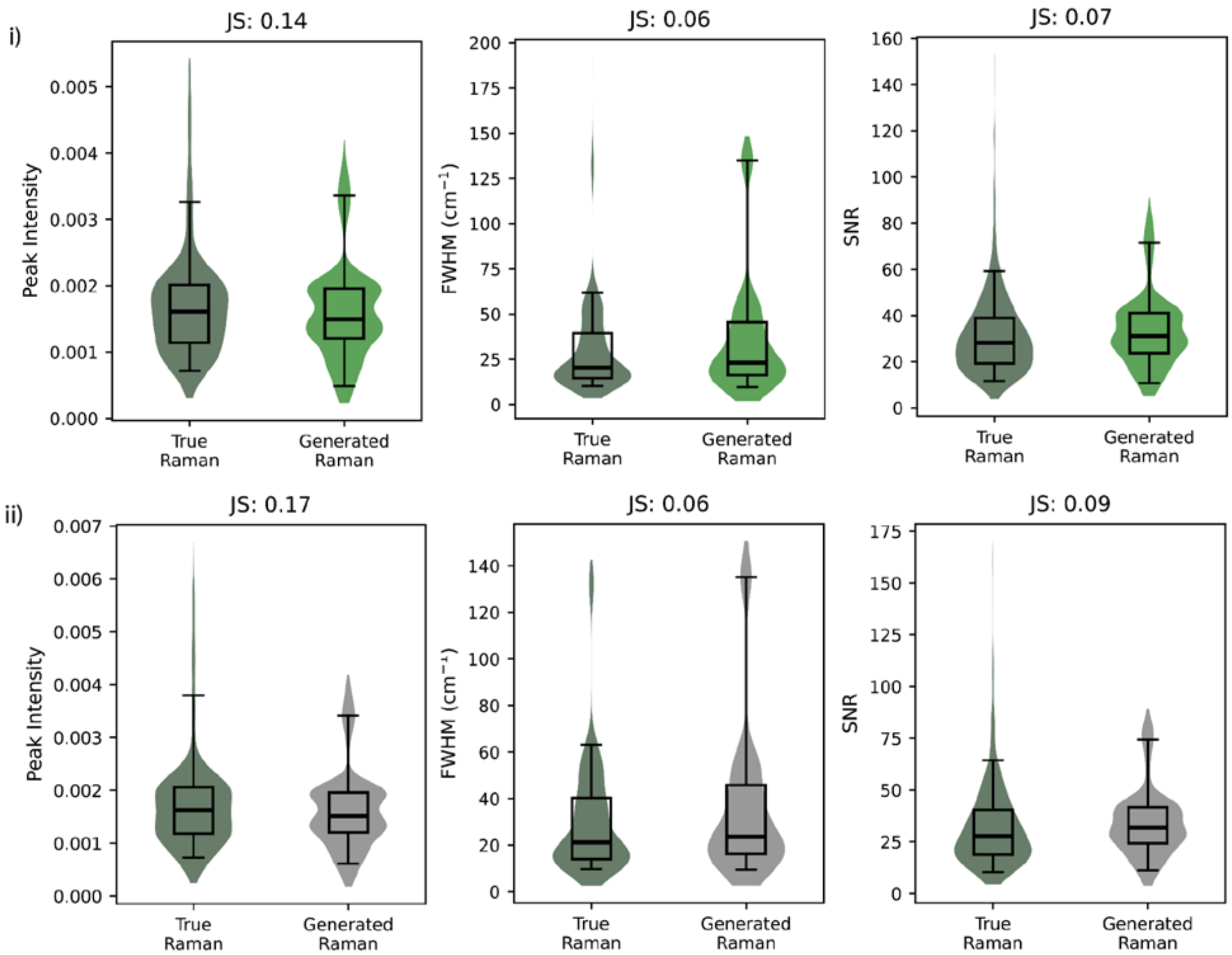

*Fig S3: Preservation of peak-level Raman features in generated spectra from GFP+ and GFP- E. coli. i, Violin plots comparing peak intensity, FWHM, and SNR between true Raman spectra (dark green) and generated spectra (fluorescent green) for GFP+ E. coli. ii, Equivalent analysis for GFP- E. coli, with generated spectra characteristics shown in grey. Box plots denote the median and interquartile range, and JS divergence values summarize distributional similarity between true and generated spectra for each feature(n= 132 GFP- cells, n=138 GFP+ cells).*

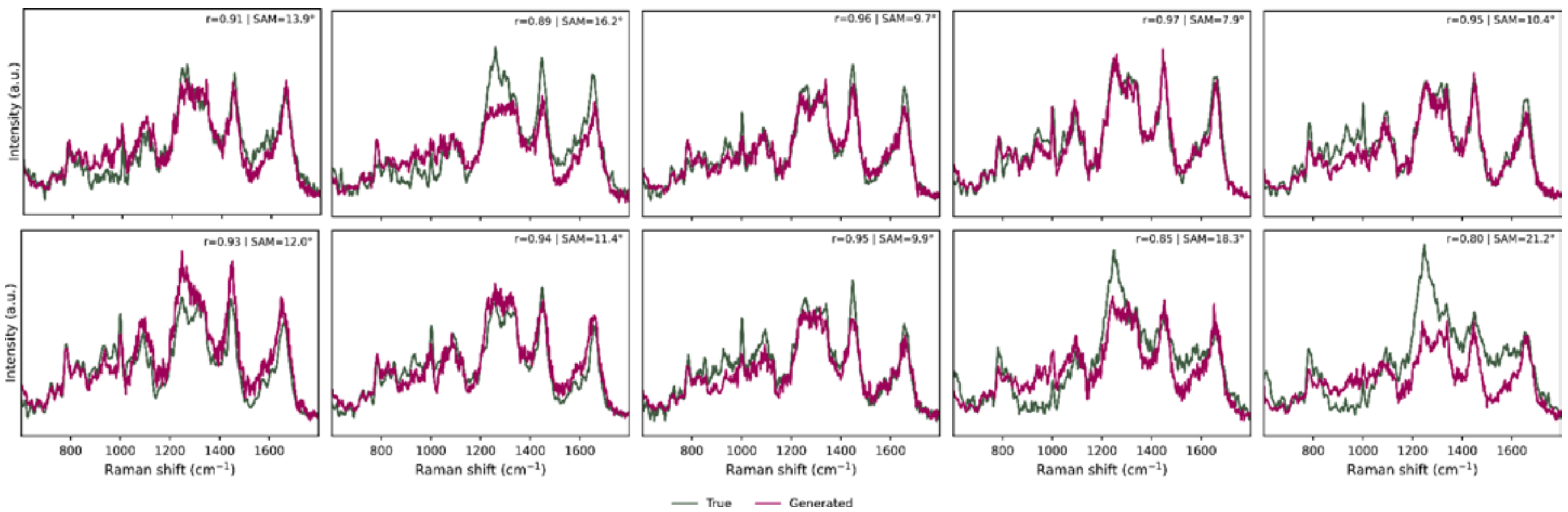

*Fig S4: Cell-resolved comparison of generated and experimentally measured Raman spectra. Representative T cell Raman spectra from 10 individual cells are shown, comparing generated spectra (pink) with experimentally measured true spectra (green). Pearson $r$ and SAM are reported for each cell to quantify spectral agreement in terms of overall shape similarity and angular deviation, respectively. Panel-to-panel variation highlights residual heterogeneity in reconstruction fidelity across cells.*

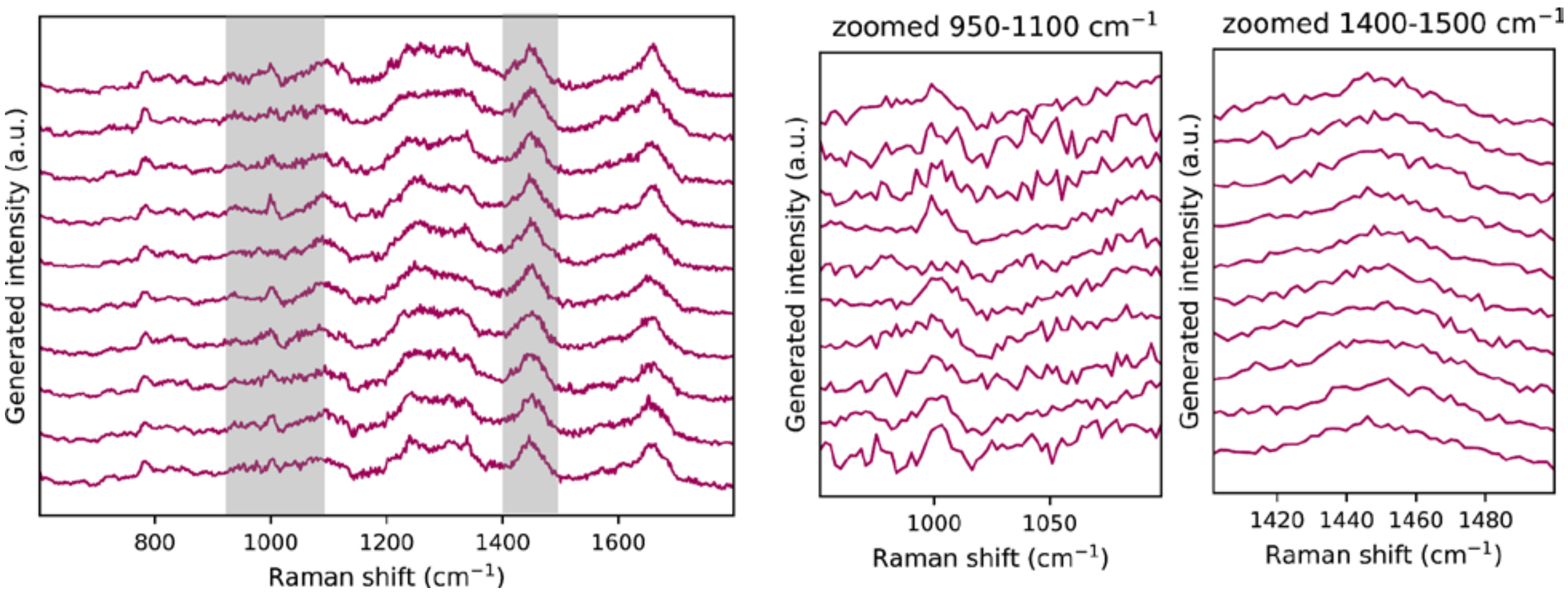

*Fig. S5. Reconstructed spectra preserve single-cell spectral heterogeneity. Stacked generated Raman spectra from n = 10 individual T cells demonstrate variability across the full spectral range and within zoomed regions spanning 950-1100 cm$^{-1}$ and 1400-1500 cm$^{-1}$. Shaded bands indicate representative intervals with clear differences in local line shape and relative peak intensity across cells.*

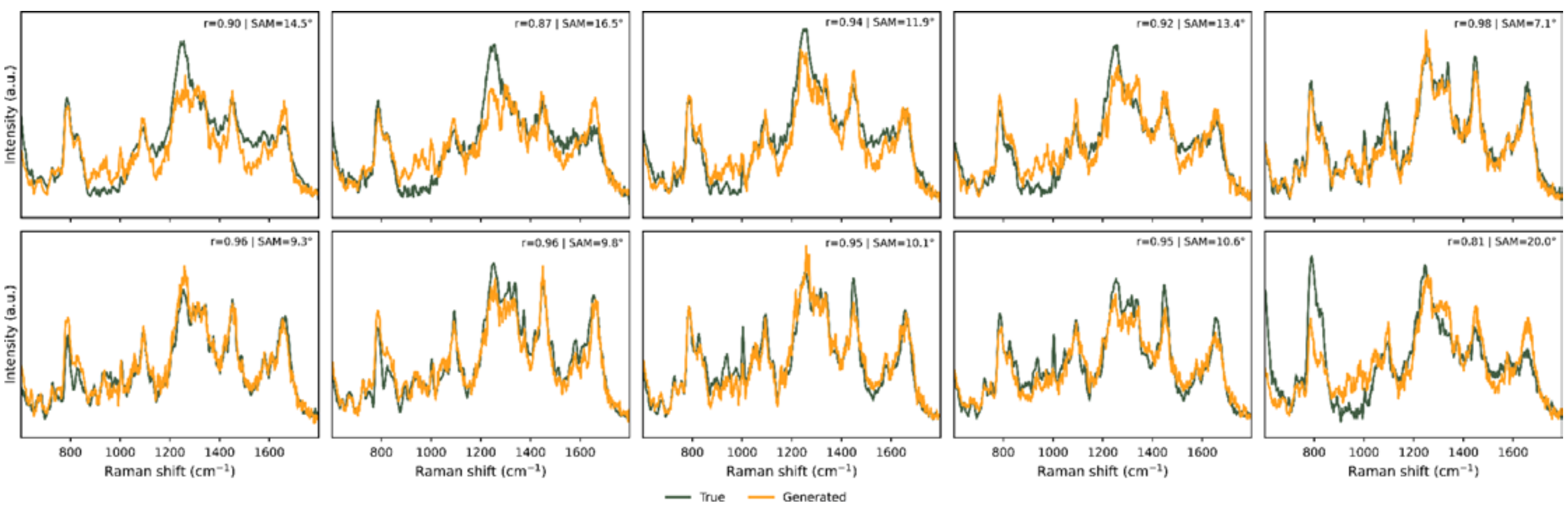

*Fig. S6. Single-cell overlays of generated and true Raman spectra for B cells. Generated Raman spectra (orange) are overlaid with experimentally acquired true spectra (green) for 10 representative individual B cells. Pearson correlation $r$ and SAM are shown in each panel to quantify spectral similarity. These cell-level comparisons illustrate preservation of dominant spectral features, including peak positions and relative band structure, while also revealing variability in reconstruction fidelity across individual cells.*

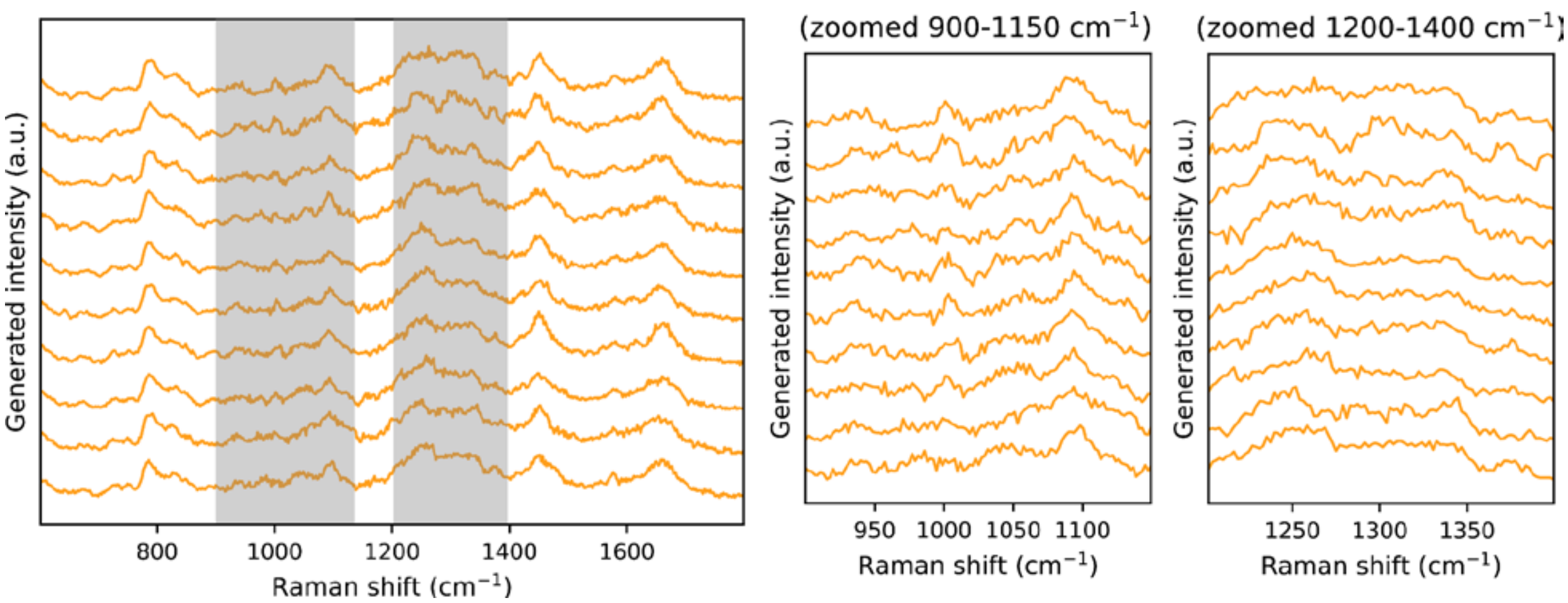

*Fig. S7. Stacked reconstructed spectra from n = 10 individual B cells, together with magnified views of two representative spectral windows, 900-1150 cm$^{-1}$and 1200-1400 cm$^{-1}$ . Grey shaded regions indicate spectral intervals with visible inter-cell variation.*

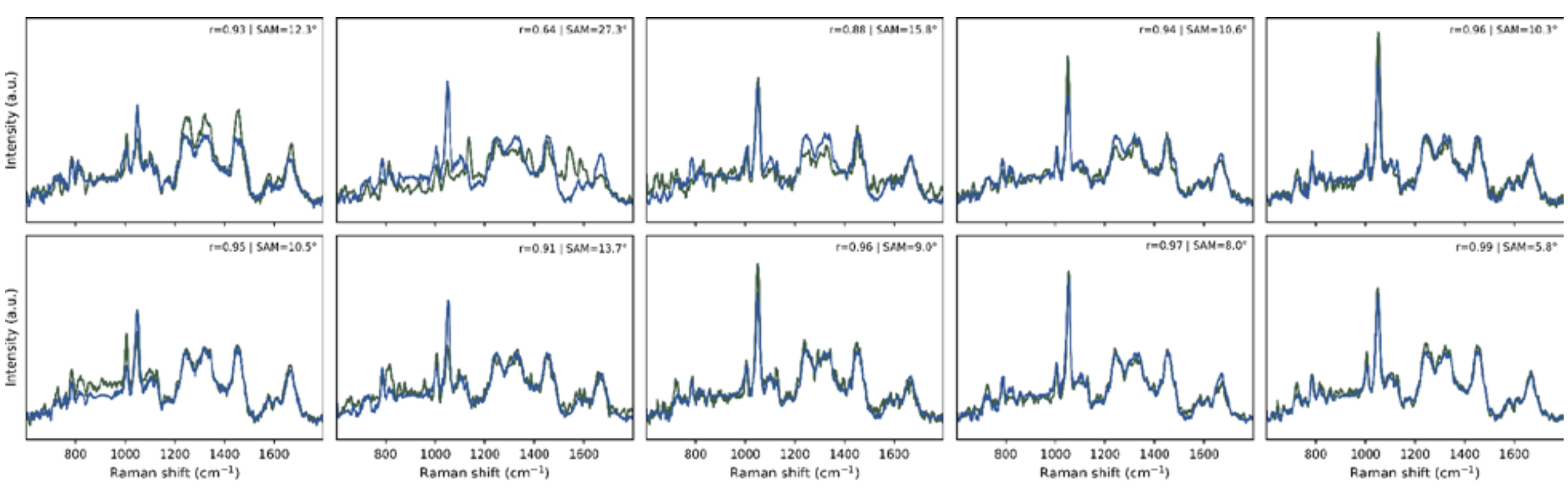

*Fig. S8. Cell-resolved agreement between generated and measured Raman spectra in bacterial cells. Representative single-cell Raman spectra from 10 bacterial cells are shown, comparing generated spectra (blue) with experimentally acquired true spectra (green). Pearson $r$ and SAM are indicated in each panel to quantify spectral agreement.*

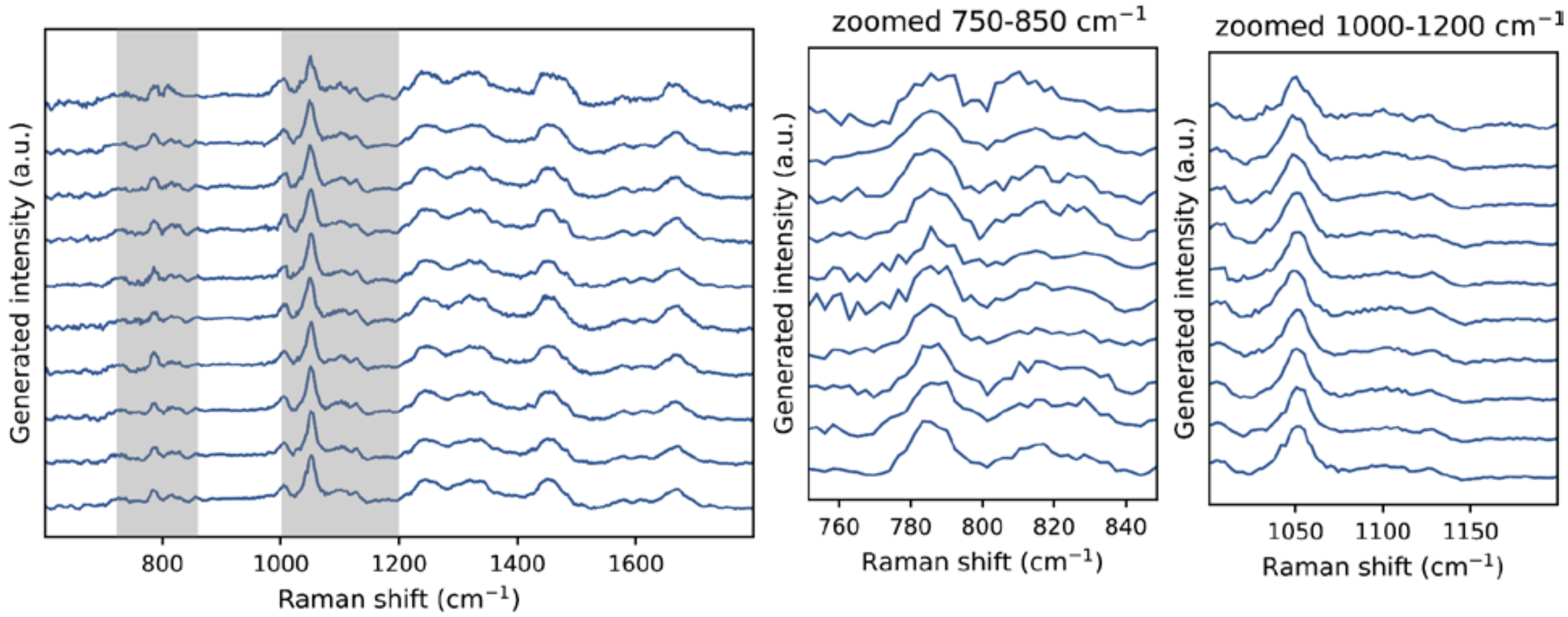

*Fig. S9. Stacked generated Raman spectra from n=10 individual bacterial cells show variability across the full spectral range and within zoomed regions spanning 750-850 cm$^{-1}$ and 1000-1200 cm$^{-1}$.*

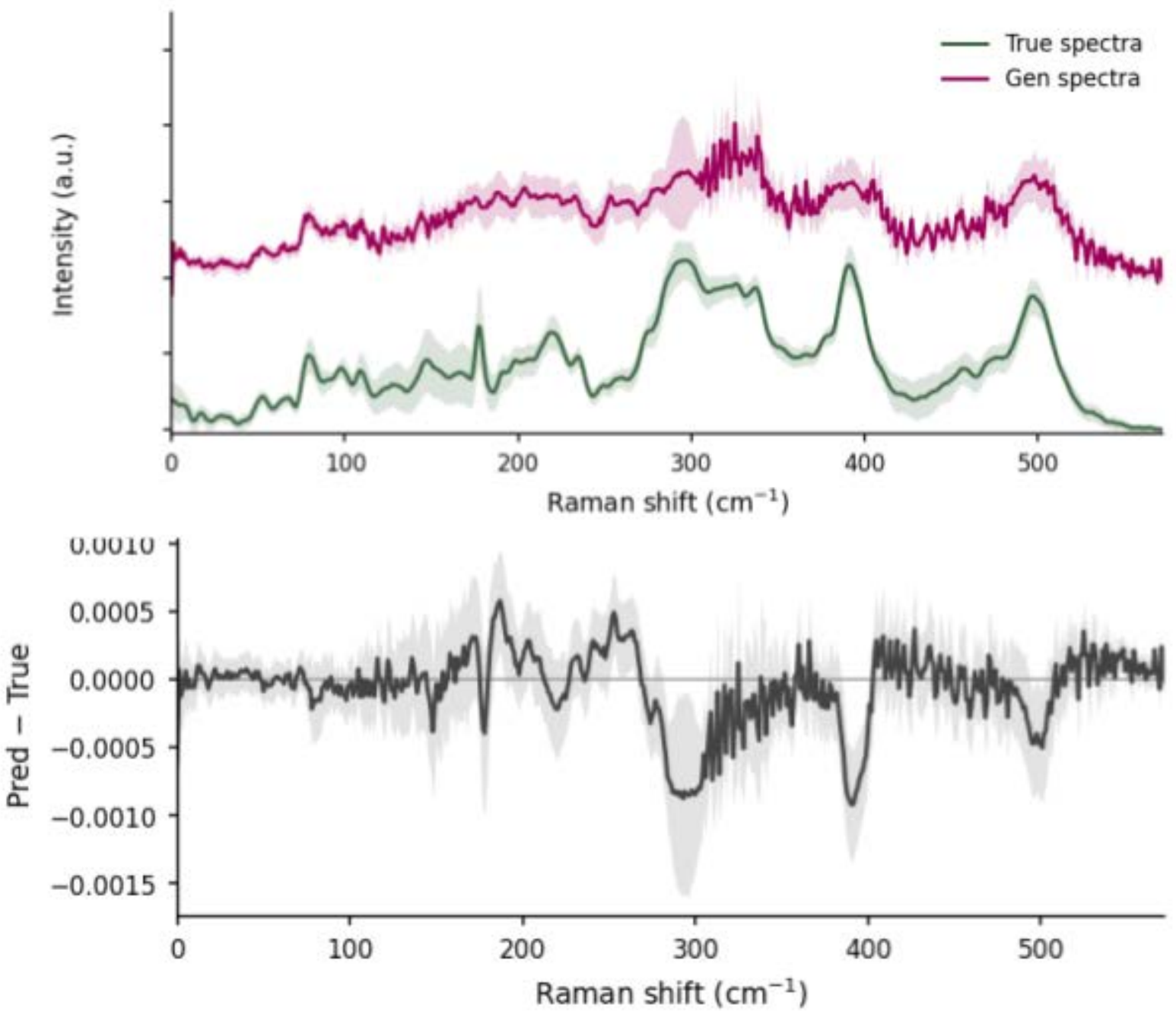

*Fig. S10. Spectral performance of the priorless dual-decoder model. Mean true Raman spectra (green) and spectra generated by the priorless dual-decoder model (pink) are shown in the top panel, with the corresponding residual spectrum (Pred − True) shown below. Structured residuals across multiple bands indicate persistent reconstruction error and imperfect alignment with the experimentally measured spectra.*

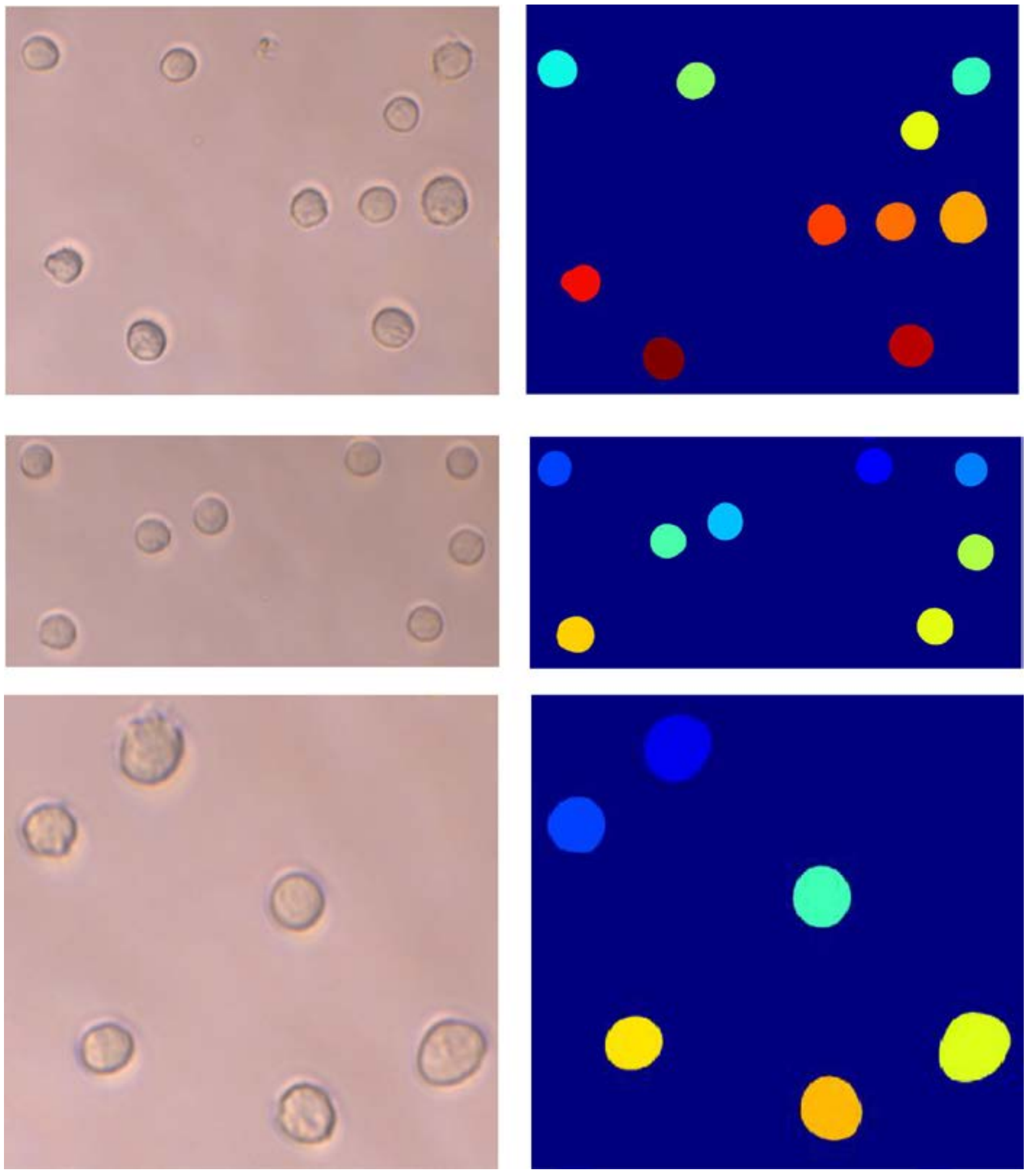

*Fig. S11. BF-based localization of individual cells for segmentation. Representative BF images (left) and corresponding localization outputs (right) used to identify cell positions before segmentation. Each colored marker denotes a detected cell instance derived from the BF image. Examples are shown across multiple representative field of views.*